\documentclass[preprint]{aastex}

\newcommand{\kms}{\mbox{km s$^{-1}$}}
\newcommand{\etal}{\mbox{et al.}}
\newcommand{\eg}{\mbox{e.g.}}

\newcommand{\Msun}{\mbox{$M_{\sun}$}}

\newcommand{\um}{\mbox{$\mu$m}} 
\newcommand{\ug}{\mbox{$\mu$G}} 
 
\newcommand{\jj}[2]{\mbox{$J = #1 \rightarrow #2$}}

\newcommand{\uco}[1]{\mbox{$^{#1}$CO}} 
\newcommand{\cuo}[1]{\mbox{C$^{#1}$O}} 
\newcommand{\blos}{\mbox{$B_{los}$}}
\newcommand{\Tbg}{\mbox{$T_{bg}$}}

\newcommand{\Tmb}{\mbox{$T_{mb}$}}
\newcommand{\Ta}{\mbox{$T_{A}^{*}$}}
\newcommand{\Tex}{\mbox{$T_{ex}$}}
\newcommand{\Tsys}{\mbox{$T_{sys}$}}
\newcommand{\tauoh}{\mbox{$\tau_{1667}$}}

\newcommand{\hii}{\mbox{\ion{H}{2}}}
\newcommand{\skipthis}[1]{}
\newcommand{\vlsr}{\mbox{$V_{lsr}$}}
\newcommand{\dV}{\mbox{$\Delta V$}}
\newcommand{\dv}{\mbox{$\Delta v$}}
\newcommand{\vm}{$\mathcal{M}$} 
\newcommand{\vw}{$\mathcal{W}$} 

\newcommand{\fmr}{($\Phi/M$)$_{\rm n}$}
\newcommand{\cm}[1]{\mbox{cm$^{#1}$}}   
\newcommand{\Nhh}{\mbox{$N$(H$_2$)}}
\newcommand{\noh}{\mbox{$N$(OH)}}

\newcommand{\hh}{\mbox{H$_2$}}
\newcommand{\power}[2]{\mbox{$#1 \times 10^{#2}$}} 
\newcommand{\hhco}{\mbox{{H$_2$CO}}}

\slugcomment{Accepted by ApJ, 27 Feb 2001}
\shorttitle{OH Zeeman survey}
\shortauthors{Bourke et al.}
  
\begin{document}

\title{New OH Zeeman measurements of magnetic field strengths in molecular
clouds}
  
\author{Tyler L. Bourke}
\affil{Harvard-Smithsonian Center for Astrophysics, 60 Garden Street MS 42, 
Cambridge MA 02138, USA; and School of Physics, University College, 
The University of New South Wales, Australian Defence Force Academy, 
Canberra, ACT 2600, Australia}
\email{tbourke@cfa.harvard.edu}

\author{Philip C. Myers}
\affil{Harvard-Smithsonian Center for Astrophysics, 60 Garden Street MS 42, 
Cambridge MA 02138, USA}
\email{pmyers@cfa.harvard.edu}

\author{Garry Robinson}
\affil{School of Physics, University College, The University of New South 
Wales, Australian Defence Force Academy, Canberra, ACT 2600, Australia}
\email{garry@ph.adfa.edu.au}

\and

\author{A. R. Hyland}
\affil{Office of the Vice Chancellor, Southern Cross University, Lismore, 
NSW 2480, Australia\altaffilmark{1}}
\altaffiltext{1}{Present Address: Deputy Vice-Chancellor, James Cook
University, Townsville, QLD 4811, Australia}

\begin{abstract}

We present the results of a new survey of 23 molecular clouds for the
Zeeman effect in OH undertaken with the ATNF Parkes 64-m radio telescope and 
the NRAO Green Bank 43-m radio telescope.  The Zeeman effect was clearly 
detected in the cloud associated with the \hii\ region
RCW~38, with a field strength of $38\pm3$ \ug, and possibly detected in 
a cloud associated with the \hii\ region RCW~57, with a field
strength of $-203\pm24$ \ug.
The remaining 21 measurements give formal upper limits to the magnetic
field strength, with typical 1$\sigma$ sensitivities $<20$ \ug.

For 22 of the molecular clouds we are also able to determine the
column density of the gas in which we have made a sensitive search for 
the Zeeman effect.
We combine these results with previous Zeeman studies of 29 molecular
clouds, most of which were compiled by Crutcher (1999), for
a comparsion of theoretical models with the data.

This comparison implies that if the clouds can be modeled as initially
spherical with uniform magnetic fields and densities that evolve to their
final equilibrium state assuming flux-freezing then the typical cloud is
magnetically supercritical, as was found by Crutcher (1999).  If the
clouds can be modeled as highly flattened sheets threaded by uniform
perpendicular fields, then the typical cloud is approximately magnetically 
critical, in agreement with Shu \etal\ (1999), but only if the true values 
of the field for the non-detections
are close to the 3$\sigma$ upper limits.  If instead these values are
significantly lower (for example, similar to the 1$\sigma$ limits),
then the typical cloud is generally magnetically supercritical.

When all observations of the Zeeman effect are considered, the
single-dish detection rate of the OH Zeeman effect is relatively low.
This result may be due to low mean field strengths, but a more
realistic explanation may be significant field structure within the
beam.  As an example, for clouds associated with \hii\ regions the
molecular gas and magnetic field may be swept up into a thin shell, which 
results in a non-uniform field geometry and measurements of the beam-averaged 
field strength which are significantly lower than the true values.  This
effect makes it more difficult to distinguish magnetically subcritical
and supercritical clouds.

\end{abstract}



\section{Introduction}

Magnetic fields are widely believed to influence the stability and
dynamics of molecular clouds and to explain the supersonic line
widths seen in all interstellar molecular spectral lines (see reviews by
Heiles \etal\ 1993; McKee \etal\ 1993; and references therein).  Yet these
central ideas in molecular cloud and star formation physics are supported
by remarkably few measurements of magnetic field strengths, due to the 
difficulty in measuring the Zeeman effect in molecular lines
such as the 18 cm lines of OH.  Excluding masers, for which the velocity
dispersion and gas density are very uncertain, there are only 17 
clouds in which the Zeeman effect has been detected 
above 3$\sigma$ (Crutcher 1999; Sarma \etal\ 2000; Crutcher \& Troland
2000).  

As discussed in
detail in the reviews by Heiles \etal\ (1993) and Crutcher (1994), the
only viable technique available for measuring magnetic field strengths in
molecular clouds is the Zeeman effect in spectral lines. 
The magnetic field reveals itself via the Zeeman
effect as small frequency shifts, $\Delta \nu_{z}$, in the right and left 
circularly polarized (RCP and LCP, respectively) components of the 
spectral line with respect to the frequency in the zero field case,
$\nu_{\mathrm{o}}$.  The magnetic field strength can be determined by
measuring this frequency shift.  However, under most astrophysical
conditions $\Delta \nu_{z} \ll \Delta \nu$, where $\Delta \nu$ is the 
full width at half maximum 
of the spectral line (the exceptions are masers, see \eg, Reid \&
Silverstein 1990), and so detecting the shift between the RCP and LCP
components due to the the Zeeman effect is difficult, and 
complete information about the magnetic field direction and magnitude is
not obtainable.  In this situation only 
the determination of the line-of-sight component of the field strength,
\blos\ (= $B\cos\theta$), and its sign (i.e., toward or away from the
observer) is possible.

The most succussful efforts have
been those of Crutcher and collaborators (e.g., Crutcher \& Kaz\`es
1983; Kaz\`es \& Crutcher 1986; Troland, Crutcher \& Kaz\`es 1986;
Crutcher, Kaz\`es \& Troland 1987; Kaz\`es \etal\ 1988, Goodman \etal\
1989; Crutcher \etal\ 1993; Roberts, Crutcher \& Troland 1995; Crutcher \etal\
1999a), mainly through observations
of the OH transitions at 1665 and 1667 MHz.  Almost all
successful detections of the Zeeman effect have used this technique,
with observations of \ion{H}{1}\ and CN also yielding success (e.g.,
Roberts \etal\ 1993; Brogan \etal\ 1999; Crutcher \etal\ 1999b).
The major telescopes that have been used are Nan\c{c}ay (OH), the Hat
Creek 85-ft (\ion{H}{1}), the NRAO 43-m (OH, \ion{H}{1}), 
Arecibo (OH, \ion{H}{1}), the VLA (OH, \ion{H}{1}), and the IRAM 30-m (CN).   

All the reliable detections, and a number of upper limits 
were compiled by Crutcher (1999) for comparison
with theory.  In addition to the field strength or upper limit for each
cloud in his analysis (27 clouds), Crutcher also compiled information on
the column density, line width, number density, temperature and 
size for each molecular cloud.  By comparing the data against
a model spherical cloud with uniform
density and uniform magnetic field, with corrections for the simple
virial terms, Crutcher concluded that (i) internal motions are
supersonic, (ii) the ratio of magnetic to thermal pressure is sufficient
for magnetic fields to be important, (iii) the mass-to-magnetic flux
ratio is about twice critical, which suggests that static magnetic
fields alone may be insufficient to provide cloud support, (iv) the
kinetic and magnetic energies are approximately equal, which implies
that MHD waves and static fields are equally important in cloud
dynamics, and (v) magnetic field strengths scale with gas densities in
agreement with the predictions of ambipolar diffusion, $|B| \propto
\rho^{0.47}$. However, we note that the scatter in this last result is 
significant, and it does not take into account measured upper limits 
to \blos.

Shu \etal\ (1999) compared the data of Crutcher (1999) with a model of a
highly flattened molecular cloud (Allen \& Shu 2000), and found that the
mass-to-magnetic flux ratio is approximately critical.  Considering the
uncertainties in the measurements and the simple nature of the model,
Shu \etal\ concluded that it could not yet be decided whether the clouds are
generally magnetically subcritical (slow evolution, dominated by ambipolar
diffusion) or supercritical (fast evolution, dominated by collapse
and fragmentation), though the distinction is
clearly very important.

Further, the role of the magnetic field in cloud support has been
questioned since magneto-hydrodynamic (MHD) simulations indicate that 
unforced MHD turbulence
decays in about a free-fall time (MacLow \etal\ 1998; Padoan \& Nordland
1999; Stone, Ostriker \& Gammie 1998), and since examination of stellar ages
indicates that star-forming molecular clouds may not require long-lived
support from either turbulence or magnetic fields (Elmegreen 2000, Hartmann
2001).

Since progress in this field clearly requires more observations,
we have undertaken OH Zeeman observations of molecular clouds in the relatively 
unexplored southern hemisphere and observations of a sample of northern
hemisphere molecular clouds not previously studied.


\section{Observations}

\subsection{Parkes}

OH Zeeman observations of southern hemisphere molecular clouds were 
undertaken in 1995 July and 1996 October
with the Australia Telescope National Facility 
(ATNF)\footnote{The Australia Telescope is funded by the
Commonwealth of Australia for operation as a National Facility managed by 
CSIRO \label{foot-atnf}} Parkes 64-m radiotelescope, 
using a HEMT receiver equipped with dual linear polarization feeds.  
This system provided cold-sky system temperatures 
of \Tsys\ $\sim$20-25 K.  The OH ground state transitions at 
1665.40184 (hereafter,
the OH 1665 line) and 1667.35901 MHz (hereafter, the OH 1667 line) were 
observed simultaneously in one bandpass of 4 MHz
centered on 1666 MHz with a channel separation of 488.28 Hz (0.087 \kms).  
The outputs from the linear
feeds were crossed in a hybrid coupler located immediately after the low
noise amplifiers of the receiver to produce right- and left-hand circular 
polarized outputs (RCP and LCP respectively).  These signals were then 
passed through a 
double-pole, double-throw polarization switch to minimize the
effects of instrument polarization and gain differences on the spectra,
switching every $\sim$10 seconds.  The
center bandpass frequency was switched by $\pm$0.25 MHz every 4
minutes.  Both senses of circular polarization were observed and
recorded simultaneously.  The FWHM beam size at these frequencies is
$\sim$13\arcmin\ and the main beam efficiency is $\sim$0.7.

To determine the sense of polarization in the output from the correlator,
OH masers having strong circular polarization were observed at the start 
of each run (Caswell \etal\ 1980; Caswell \& Haynes 1983, 1987a).  
We used Zeeman
observations of the deep OH absorption line toward Orion~B to test our
setup.  In both observing periods we were able to reproduce the results
of Crutcher \& Kaz\`es (1983; hereafter CK83) after $\sim$5 hr
integration time.  By combining the 1665 and 1667 OH
results from 1995 and 1996 a field strength of $35\pm2$ \ug\ is inferred
for the subcomponent
identified by CK83 as the source of the Zeeman pattern, essentially
independent of hour angle.  The Stokes $I$ and $V$ spectra for Orion~B 
observed in 1996 are shown in Figure~\ref{fig-orionb}.
These values are in agreement with the value determined by CK83 
($38\pm1$ \ug) using the Nan\c{c}ay telescope.  It should be noted
that the magnetic field strength inferred for Orion B is not unique, 
but depends on 
the properties (line strength, line width) of the subcomponent, which
are not fully constrained by the observations.

The displacement of the right- and left-circularly polarized beams on
the sky, commonly called ``beam-squint" (e.g., Troland \& Heiles 1982) can 
result in a spurious Zeeman effect if there exists a sufficiently large
component of velocity gradient in the
spectral line along the same direction as the squint.  This effect is
diminished considerably with the combination of an altitude-azimuth telescope 
such as Parkes and a long on-source integration time, since the source then
rotates within the beam and the true Zeeman signal is modulated.  For Parkes 
we measured a squint of $\sim$15\arcsec\
for both observing runs, but with different position angles, most
likely caused by the replacement of the focus cabin at the end of 1995.
While this is a relatively large value for the squint, the fact that we
were able to reproduce the results for Orion~B both times with the same
derived magnetic field strength suggests that the squint is not
a problem for our observations.  If a spurious Zeeman effect had been caused 
by squint then the inferred magnetic field strength for the OH 1665 
line would be 1.7 times greater than the OH 1667 line, which we do
not see in our results.

\subsection{Green Bank}

OH Zeeman observations of northern hemisphere molecular clouds were
undertaken in 1996 September and 1998 June with the National Radio
Astronomy Observatory 
(NRAO)\footnote{The National Radio Astronomy 
Observatory is a facility of the National Science Foundation operated
under cooperative agreement by Associated Universities, Inc.}
Green Bank 43-m telescope, using a 
dual channel cooled HEMT receiver with circular polarization feeds.
Cold sky system temperatures of 20--25 K were recorded with this system.
The OH 1665 and 1667 MHz lines were observed simultaneously in both
polarizations with frequency switching with a channel separation of 1220 Hz
($\sim$0.22 \kms).  To minimize the effect of instrument polarization
and gain differences on the spectra polarization switching was performed
at a rate of $\sim$1 Hz.  The setup is similar to the one used by
Crutcher \etal\ (1993).  The FWHM beam size at these frequencies is
$\sim$18\arcmin\ and the main beam efficiency is $\sim$0.7.

As was the case for the Parkes observations, we tested the setup with
OH Zeeman observations of Orion~B.  The Zeeman effect was clearly seen
during both observing sessions, and by combining the
1665 and 1667 OH results for 1996 and 1998 we infer a magnetic field 
strengths of $34\pm3$ \ug\ for the subcomponent identified by CK83. 

We measured the beam squint of the 43-m telescope to be no more than 1\arcsec. 
Beam squint is more important for an equatorially mounted telescope such as
the 43-m compared to an alt-az telescope, if the squint happens to align
with a velocity gradient within the source being observed.  
However, the chances of such an alignment are small, and the change in hour
angle will not cause any change in alignment between the squint and velocity
gradient.  

\subsection{Mopra Observations}

As OH and \uco{13}\ are believed to trace similar column densities
(\eg, Myers \etal\ 1978; Heiles \etal\ 1993), we have mapped a number of
the southern clouds (those observed at Parkes) in the \uco{13}\ \jj{1}{0}
transition, 
in order to provide an independent measurement of the column density,
and to resolve more structure than is possible with the poor angular
resolution provided by the OH observations.
The observations were undertaken in April 1997 with the 22-m ATNF Mopra
radiotelescope.  At the \uco{13}\ \jj{1}{0}\ frequency of 110201.37 MHz
only the inner 15-m of the surface was illuminated at the time of our
observations, providing an effective beam size of 45\arcsec.  
The receiver was a dual linear polarization SIS receiver 
with a total system temperature measured
to be $\sim$150 K in good weather.  The 16384 channel autocorrelator
was configured for two IFs, providing a spacing between channels of 0.17
\kms\ over a velocity range of $\sim$220 \kms\ (1024 spectral channels
with a 64 MHz bandwidth) at the \uco{13}\ \jj{1}{0}\ frequency.  One IF
was centered on the \uco{13}\ \jj{1}{0}\ frequency, the second being
centered on the SiO $\nu$ = 1, \jj{2}{1}\ transition at 86243.44 MHz for
regular pointing checks using SiO masers listed in the SEST Handbook
(\url{http://www.ls.eso.org/lasilla/Telescopes/SEST/SEST.html}).
  
The mapping was performed on a 1\arcmin\ grid
in position switching mode, with an on-source integration time of 4
minutes per position.  
Removal of first-order baselines was
sufficient in most cases to correct for any system effects not removed
by the reference observation.  The spectra were calibrated
offline for atmospheric attenuation and the variation of telescope beam
efficiency with elevation, and the intensity scaled to the \Ta\ scale of
SEST by observations of Orion ($\alpha_{1950} = 05^{\rm h}32^{\rm m}47\fs0$,
$\delta_{1950} = -05\arcdeg24\arcmin23\arcsec$) in the \uco{13}\ \jj{1}{0}\ 
transition, assuming \Ta(Orion) = 13 K (Appendix of the SEST Handbook).  
The conversion to \Tmb\ is given by \Tmb\ = 
\Ta/$\eta_{mb}^{\mathrm SEST}$ where $\eta_{mb}^{\mathrm SEST}$ = 0.7
near 110 GHz.  In general we attempted to map down to at least
the 50\% level of the peak emission in each source.

\section{Source Selection}

\subsection{Southern sources}

Our initial search for suitable
sources centered on the OH studies of Robinson, Goss \& Manchester (1970), 
Goss, Manchester \& Robinson (1970), Manchester, Robinson \& Goss (1970), 
Robinson, Caswell \& Goss (1971), Caswell
\& Robinson (1974), and Turner (1979).  The spectral resolution of
these previous observations, many undertaken with Parkes, was generally
insufficient to detect weak, narrow maser features, and so we reobserved
all our potential targets with the setup described here.  We also
reobserved a number of other sources from these lists with a greater
velocity coverage, as well as some sources from Dickel \& Wall (1974),
Toriseva, H\"oglund \& Mattila (1985), and Chan, Henning \& Schreyer (1996).  
None of the sources we surveyed from these lists showed any significant 
OH absorption that was not previously known.  

In order to select suitable Zeeman candidates we required observations
with substantial improvements in sensitivity and spectral resolution over
the earlier OH surveys of southern HII regions.  We thus re-observed 
$\sim$80 sources drawn from these
lists to assess their suitability.  To determine
our primary candidates we used the ``sensitivity estimator" given by 
Troland (1990; see also Goodman 1989) to estimate the integration time 
required per 
source to reach our target sensitivity of 3$\sigma$ $\sim$30 \ug.
In order to have the greatest sensitivity to the Zeeman effect, the
spectral line profiles of the selected sources should be narrow and
strong against a weak continuum source (so that the system temperature
is as low as possible), and free of maser emission.
Using these criteria we chose the 9 sources for our primary candidate
list, selecting those sources with the lowest integration times required
to reach the target sensitivity.

Previously unreported masers discovered in the survey are discussed in
Appendix~\ref{appen-masers}.

\subsection{Northern sources}

Crutcher and co-workers have previously observed the most promising
OH Zeeman candidates in the northern sky (Crutcher 1999).  However, a
number of promising sources remain to be studied.  The best list to date
is the extensive survey of OH by Turner (1979) using the NRAO 43-m telescope.
Two of the previous detections of the Zeeman effect, S88B 
(Crutcher \etal\ 1987) and S106 (Kaz\`es \etal\ 1988),
have only modest OH line absorption depths of \Ta\ $\sim$ 0.5 as observed 
with the 43-m,
but they are among the few sources with clearly detected Zeeman
patterns.  The Turner catalog contains a number of sources with similar
line absorption depths.
To select sources for Zeeman observations with the 43-m, we re-observed
a number of candidates from the Turner catalog, as well as sources
identified as \hii\ regions from the IRAS database, sensitive continuum
surveys, and water maser surveys (e.g., Chan \etal\ 1996; Carpenter,
Snell \& Schoerb 1990; Wouterloot, Brand \& Fiegle 1993; Wouterloot \&
Brand 1989).  We observed $\sim$200 sources, and
identified 11 Zeeman candidates using the same criteria as for the
Parkes sample.

The final source list is given in Table~\ref{tbl-obs}.  Column 1 lists
the source name, columns 2--3 the source position, and columns 4--6 the
results of fitting a Gaussian to the OH 1667 line profile.  Column 7
lists the continuum source temperature, where appropriate, and column 8
list the total on-source integration time for the Zeeman observations.


\section{Results \& Analysis}

In Zeeman experiments, the magnetic field reveals itself as frequency 
shifts in the right and left circular polarized components ($I_{RCP}$ and 
$I_{LCP}$, respectively) of the spectral line.  A full treatment of the
Zeeman effect under astrophysical conditions may be found in Sault \etal\
(1990; see also Heiles \etal\ 1993), and we include only a brief discussion 
of the analysis here.  
In the situation where the frequency shift is much smaller than the line width
(which is always the case for non-maser lines), the difference or Stokes $V$ 
spectrum, $V = I_{RCP} - I_{LCP}$, is only sensitive to
the line-of-sight component of the magnetic field and can be modeled as
\begin{equation}
V = \frac{b}{2}B{\rm cos} \theta \frac{dI}{d\nu} + \beta I ~~,
\label{eqn-zee2}
\end{equation}
where $b$ is the ``Zeeman factor" and equals 3.27 and 1.96 Hz
$\mu$G$^{-1}$ respectively for the 1665 and 1667 MHz lines of OH, $B$ is
the magnetic field strength,
$\theta$ is the angle between the field lines and the
line-of-sight, $\beta$ is the ``gain term'' arising from gain
differences between the RCP and LCP signal paths (Sault \etal\ 1990) and 
$I$ is the Stokes $I$ spectrum (=$I_{RCP} + I_{LCP}$). 
The Zeeman term in equation~(\ref{eqn-zee2}), 
$(b/2)B {\rm cos} \theta (dI/d\nu)$, 
reveals itself in the $V$ spectrum as the characteristic sideways ``S", or 
``Zeeman pattern".  The line-of-sight field strength 
$\blos = B{\rm cos} \theta$ is inferred 
from equation~(\ref{eqn-zee2}) by least-squares fitting the right hand side of 
this equation to the observed $V$ spectrum. 
Other fitting techniques are discussed and analysed in Sault \etal\ (1990).

\subsection{Fitting $dI/d\nu$ to $V$}

In Table~\ref{tbl-zee1} we present the results of our Zeeman analysis.  
In this Table column 1 lists the source name, column 2 the velocity
of the OH feature, column 3 the magnetic field strength determined
from the OH 1665 transition with its one sigma uncertainty, column
4 as for column 3 but for the OH 1667 transition, and
column 5 the weighted mean line-of-sight field strength $B_{los}$ 
(found by combining the results for the 1665 \& 1667 lines) and its weighted 
one sigma uncertainty $\sigma_B$ obtained by standard error analysis.  

In the majority of sources the line profiles are simple and well fitted
by a single Gaussian profile.  Notable exceptions are RCW~38, RCW~57,
W7, and G10.2-0.3.  In these cases, alternative approaches to 
performing the Zeeman analysis are described in CK83, or Heiles (1988) and 
Goodman \& Heiles (1994).  In the method of CK83, the $V$ spectrum
shows a clear ``S'' structure signifying the Zeeman effect, but it
is too narrow to arise from the entire $I$ line profile.  By integrating the $V$
spectrum and fitting the result with a Gaussian it is possible to infer
the properties (\vlsr, \dV) of the component of the $I$ spectrum
responsible for the ``S''.  The reality of such a component is
strengthened if the same properties are derived for both the OH 1665 and 1667
lines.  CK83 applied this method to Orion~B (see Fig.~\ref{fig-orionb}), 
and we have
successfully used this technique to analyse RCW~38 (see below).
In Heiles (1988) and Goodman \& Heiles (1994) the 
$I$ profile is broken down into independent Gaussian
subcomponents, and the $V$ spectrum is then modeled as the sum of
derivatives of these components, with unique values of \blos\ assigned to 
each component.  We have applied this method to those sources for which
a single Gaussian could not adequately be fitted to the $I$ spectrum 
and where the CK83 method is inappropriate, because there is no clear
``S'' in the $V$ spectra.  These are indicated in the notes to 
Table~\ref{tbl-zee1}.  For G10.2--0.3, we were unable to fit a
consistent set of Gaussians to the OH 1665 and 1667 lines.

In order for the derived magnetic field strength \blos\ to be considered
as a detection we require that the field strengths measured in
both OH lines be consistent with each other, 
$\mid B_{1665} - B_{1667} \mid$ $<$ $\sigma_B$, and the
field strength be greater than 3$\sigma_B$, $\mid B_{los} \mid$ $>$
3$\sigma_B$.  Only 2 sources, RCW~38 and RCW~57, can be considered as
detections using these criteria.  These are discussed below.
We consider the remaining 21 measurements to give upper limits to \blos\
averaged over our 13\arcmin\ (Parkes) and 18\arcmin\ (Green Bank) beams
(see also \S\ref{sec-virial}).  

Details of other properties for the individual sources, such as column 
densities, are given in Appendix~\ref{appen-sources} 
(see also Table~\ref{tbl-models}).

\subsection{Determination of Column Densities}

The intensity of a spectral line from a uniform medium in front of a
background region may be expressed as
\begin{equation}
T_{mb} = f [J(T_{ex})-J(T_{C})][1-\exp(-\tau)] ~~,
\end{equation}
where $T_{mb}$ is the main-beam brightness temperature, $\tau$ is the
optical depth of the cloud in the molecular transition,
$f$ is the beam filling factor, $T_{ex}$ is the excitation temperature of the
transition, $T_{C}$ is the area-weighted mean background temperature 
(the cosmic background temperature \Tbg\ plus any contribution from a 
background continuum source, $T_S$), and
$J(T) = (h\nu/k)/(\exp(h\nu/kT)-1)$ is the Planck function.
For sources observed in absorption against a compact continuum source,
$f$ represents the fraction of the beam area covered by the continuum source.  
At the frequencies of the OH ground state transitions near 1.6 GHz $J(T)
\equiv T$ for the temperatures observed in molecular clouds ($T \lesssim
100$ K), and so
\begin{equation}
T_{mb} = f [T_{ex}-T_{C}][1-\exp(-\tau)] ~~.
\label{eqn-radtrans}
\end{equation}
The observed antenna temperature \Ta\ is proportional to \Tmb, with a
proportionality constant $\eta_{mb}$ (the main beam efficiency).
With this knowledge it is possible to determine the column density of
OH, $N(\mathrm{OH})$, which is given by (Goss 1968; Turner \& Heiles
1971)
\begin{equation}
N(\mathrm{OH}) = \frac{8\pi k\nu}{hc^2A} \Tex\ \frac{\sum g_i}{g_u}
\int \tau d\nu ~~,
\end{equation}
where $\nu$ is the frequency of the transition and $A$ is its Einstein A
coefficient, $g_i$ is the statistical weight for level $i$ and $g_u$ is
the statistical weight for the upper state of the transition.  Thus, for
the OH 1667 line
\begin{equation}
N(\mathrm{OH}) = 2.25 \times 10^{14}
\left(\frac{\Tex(1667)}{\mathrm{K}}\right) \tau_{1667}
\left(\frac{\dv}{\kms}\right) \cm{-2}\  \, \, .
\label{eqn-oh-col}
\end{equation}

For clouds unassociated with \hii\ regions, previous observations have
shown that \Tex\ is generally about 6 K (e.g., Turner 1973; Crutcher
1977, 1979), and we assume this value here.  Clouds associated with 
\hii\ regions are warmer, and so we assume \Tex\ = 10 K for those
clouds.  For a cloud seen in absorption against a typical \hii\ region 
$T_C \gg \Tex$ and \Tex\ drops out of equation~(\ref{eqn-radtrans}).  We
measure the product $fT_C$ at the telescope as \Tsys(on-source) --
\Tsys(off-source), and the optical depth is then determined from 
equation~(\ref{eqn-radtrans}), and the column density from 
equation~(\ref{eqn-oh-col}).  
To determine the molecular hydrogen column density
we assume a standard OH abundance ratio of [H]/[OH] = \power{2.5}{7}\
(Crutcher 1979), which is uncertain by about a factor two.
High-angular resolution observations indicate that 
the OH abundance in unshocked gas found in PDRs associated with \hii\
regions could be enhanced by as much as a factor of 5 over this value 
(Roberts et al.\ 1995; Sarma et al.\ 2000).  
Comparison of the data for NGC~6334 in Table~\ref{tbl-models} suggests that
this is not a concern for the single-dish observations.  Sarma et al.\
(2000) use an OH abundance five times greater than the dark cloud value
and determine \Nhh\ $\sim$ \power{5}{22}\ \cm{-2}, whereas we find
\Nhh\ $\sim$ \power{2}{22}\ \cm{-2}.  The discrepancy would be greater
if we used the same abundance as Sarma \etal\ (2000).  Further
support for the OH abundance ratio we use is given below.

Where it is possible to estimate the \uco{13}\ column
density independently, or it has been previously determined 
(see appendix for individual cases), the values of the molecular hydrogen 
column density derived independently from OH and \uco{13}\ generally agree 
to within a factor two.
For sources where a map of the \uco{13}\ emission exist, the line profile of
\uco{13}\ averaged over the map is remarkable similar to OH, with similar
line velocities (Fig.~\ref{fig-rcw57-13co-oh}).  The agreement in 
line profile and the
molecular hydrogen column density derived from \uco{13}\ and OH
observations supports the assumption
that these molecules trace similar physical conditions and 
the values for \Tex\ and [H]/[OH] used here.

\subsection{RCW~38}
\label{sec-rcw38}

The Stokes $I$ and $V$ spectra for RCW~38 are shown in Figure~\ref{fig-rcw38}.  
The characteristic ``S'' shape of the Zeeman effect is clearly evident in the
$V$ spectra, but is not well fitted by a scaled version of $dI/d\nu$.
As discussed above, this suggests that the observed Zeeman profile is not 
due to the entire $I$ profile.  
The Zeeman profile is due to a subcomponent of the observed line, 
and we integrated the $V$ spectra to determine its properties.  
The integrated $V$ spectra are well fitted by single Gaussians,
essentially with identical properties for the OH 1665 and 1667 lines.
The line velocity and width for this component is given Table~\ref{tbl-zee1}.
From the fit of the derivative of this Gaussian to the OH lines we infer
a value for the magnetic field of \blos\ = $38\pm3$ \ug.

RCW~38 was observed during both observing sessions at Parkes.  The
results are consistent with each other.  This helps to validate our
result as a true Zeeman detection, as the direction of beam squint for the
two observing sessions was different, even though the magnitude of the
squint was large for Zeeman observations.  In addition, no velocity
gradients are apparent in a map of the OH absorption made during the
1996 observations.  Corroborating evidence for the reality of the RCW~38
Zeeman detection comes from high angular resolution interferometric
observations of the OH absorption made with the ATNF Compact Array
(Bourke \etal\ 2001 in prep.).  These observations indicate a compact, 
spatially isolated absorption feature with line velocity and width similar to 
those of the sub-component we infer as giving rise to the Zeeman effect.

RCW~38 is a bright \hii\ region which has recently been
observed in the near-infrared with the VLT (Alves \etal\ 2001 in prep).  In a
separate paper we present a detailed molecular line and continuum study
of the region (Bourke \etal\ 2001 in prep.) which indicates that the Zeeman
effect arises from a molecular clump on the western edge of the \hii\
region, possibly associated with the bright 10 \micron\ peak IRS1
(Smith \etal\ 1999).  Assuming \Tex\ = 10 K we determine \tauoh\ = 0.15, 
\noh\ = \power{7.6}{14}\ \cm{-2}, and \Nhh\ = \power{9.3}{21}\ \cm{-2}.  
\uco{13}\ \jj{1}{0}\ observations imply \Nhh\ = 
\power{4}{21}\ \cm{-2}\ (assuming [\hh]/[\uco{13}] = \power{7}{5}), lower 
than derived from the OH observations, which may
suggest that the OH is sampling denser gas (\eg, Roberts \etal\ 1995
for S106).  Observations of CS
\jj{2}{1}\ and \cuo{18}\ \jj{1}{0}\ support this view (Bourke \etal\ 2001
in prep.).    

\subsection{RCW~57}
\label{sec-rcw57}

The Stokes $I$ and $V$ spectra for RCW~57 are shown in 
Figure~\ref{fig-rcw57-zee}.  
The $I$ profile is well fitted by two overlapping Gaussians, a
deep absorption component at \vlsr\ = --26.1 \kms\ and a weaker component at
\vlsr\ = --22.6 \kms.  The \uco{13}\ \jj{1}{0}\ channel maps shown in
Figure~\ref{fig-rcw57-13co} reveal two spatially distinct clouds.  The
cloud in the north-east corresponds in velocity to the blue-shifted OH
component, and the larger cloud to the south-west to the red-shifted OH
component, which is the component for which we claim the detection of
the Zeeman effect.  The sum of \uco{13}\ emission spectra over the mapped region
is remarkably similar to the OH absorption spectra
(Fig.~\ref{fig-rcw57-13co-oh}), and is well
fitted by two Gaussians with similar line velocities and widths as those
fitted to the OH spectra.

The formal results for RCW~57 indicate a solid
detection of the Zeeman effect in the $-22.6$ \kms\ component, with
\blos\ = $-203\pm24$ \ug.  However, inspection of the $V$ spectra in
Figure~\ref{fig-rcw57-zee} raises some doubt about this result.
The blending of the two components makes it difficult to say with
certainty that the Zeeman effect has been detected.  At velocities
greater than about $-22.2$ \kms\ there is no overlap between the OH
velocity components, and the Zeeman effect is clearly seen, but only
part of the ``S'' is evident, due to the line overlap at lower
velocities.  As the \uco{13}\ observations show two spatially distinct 
clouds, follow-up OH Zeeman observations with the ATNF Compact Array should 
be able to resolve this question.

Located at a distance of 3.6 kpc (kinematic, \eg, Caswell \& Haynes
1987b) or 2.4 kpc (Persi \etal\
1994, based on the possible association of HD 97499 with the molecular
cloud), RCW~57 (NGC 3576) is an \hii\ region with ongoing star formation
(\eg, Persi \etal\ 1994).  The few published studies of this region 
suggest it is a ``typical" \hii\ region with associated molecular clouds.  
Whiteoak \& Gardner (1974), observed the 5 GHz transition of \hhco\ in 
absorption toward RCW~57 with a 4\farcm2 beam, and found two velocity 
components with almost identical 
properties to the Gaussian components we fitted to our OH data, 
suggesting that at
least part of the OH absorption arises in somewhat denser gas 
(10$^4$ cm$^{-3}$) than is usually sampled by OH (10$^3$ cm$^{-3}$).   
For the component at --22.6 \kms\ we find \tauoh\ = 0.06 and \noh\ =
\power{8.9}{14}\ \cm{-2}, or \Nhh\ = \power{1.1}{22}\ \cm{-2}\ compared
with \Nhh\ = \power{1.3}{22}\ \cm{-2}\ from \uco{13}\ Mopra observations.
This is the component where we have a probable detection of the Zeeman
effect, with \blos\ = $-203\pm24$ \ug.
For the component at --26.1 \kms\ we find \tauoh\ = 0.2 and \noh\ =
\power{1.5}{15}\ \cm{-2}, or \Nhh\ = \power{1.9}{22}\ \cm{-2}\ compared
with \Nhh\ = \power{1.5}{22}\ \cm{-2}\ from \uco{13}\ Mopra observations.

\section{Discussion}

Of the 23 Gaussian components we could identify along the line-of-sight
toward the 20 sources observed, the Zeeman effect was clearly detected
in one source (RCW~38) and possibly in one component of another (RCW~57),
as indicated in Table~\ref{tbl-zee1}.  In Table~\ref{tbl-zee1} we label 
our 23 components as those directly associated with \hii\
regions (those clouds whose OH absorption line velocities are similar to
the recombination line velocity of the background \hii\ region) and those 
associated with (cold) molecular clouds along the line-of-sight.  We find 
that 2 of 13 components
associated with \hii\ regions are detected, while none of the 10
components associated with cold molecular clouds are detected.  The
median 1$\sigma$ sensitivities for the \hii\ and cold-cloud samples are
respectively 11 and 9 \ug.

A comparison of OH optical depth against line width shows no clear
trend, nor does a comparison of integrated OH line area against optical
depth.  The molecular clouds showing Zeeman detections do not have
properties clearly different to the other clouds in our sample.

\subsection{The magnetic field and virial equilibrium}
\label{sec-virial}

If the gas is well coupled to the magnetic field, then the large scale
magnetic field can provide support against gravitational collapse.  
An important parameter in the discussion of support from large scale
magnetic fields is the magnetic flux-to-mass ratio, $\Phi/M$.  
To examine whether
this large scale field alone can provide support to molecular clouds,
we ignore terms in the virial theorem other than the magnetic
(\vm) and gravitational (\vw) energies, and assume that these energies
are in equilibrium.  This allows us to determine a critical mass,
$M_{\rm cr}$, when gravitational forces are balanced by magnetic 
stresses (this
has been discussed in detail in McKee \etal\ (1993); see also McKee
1999).  It is found that \vm\ = \vw\ implies
\begin{equation}
M_{\rm cr} = \frac{c_{\Phi} \Phi}{\sqrt{G}}
\end{equation}
where $G$ is the gravitational constant, and $c_{\Phi}$ is a numerical
factor which depends on the geometry under consideration.
For a spherical ``parent" cloud with uniform density and uniform magnetic 
field, $c_{\Phi}$ $\approx$ 0.12 (Mouschovias \& Spitzer 1976; Tomisaka,
Ikeuchi \& Nakamura 1988) for the final equilibrium configuration, which 
is flattened along the field lines (assuming flux-freezing).  
If the cloud is an isothermal disk (i.e., sheet-like) with a constant 
flux-to-mass ratio (uniform field and uniform column density), 
then $c_{\Phi}$ = 1/2$\pi \approx 0.16$ (Nakano \& Nakamura 1978; Shu \etal\
1999).

The critical mass can be related to observable quantities with
measurements of the magnetic field strength $B$ and column density
$N$ (McKee \etal\ 1993; Crutcher 1999), so that
\begin{equation}
\frac{\Phi}{M} = \frac{B}{mN}
\label{eqn-massflux}
\end{equation}
where $m$ is the mean particle mass (2.33$m_{\rm H}$ for 10\% He).  
The column density in equation~(\ref{eqn-massflux}) 
is the total column density, $N = 1.2\Nhh$, for
10\% He.  The magnetic field strength is the total field strength, whereas
Zeeman experiments where the Zeeman shift is less than a line width are only 
sensitive to the field along the
line-of-sight, and hence measure \blos.  As discussed in Crutcher
(1999), we can statistically estimate the line-of-sight field
strength for a large ensemble of measurements of uniform fields
oriented randomly with-respect-to the line-of-sight, with the result
that $<\blos>$ = $\mid B \mid$/2 and $<\blos^2>$ = $\mid B \mid$$^2$/3. 
In addition, for the sheet-like geometry a correction factor to the
observed column density is required, again due to the orientation
with-respect-to the line-of-sight.  In particular, if the magnetic field is
preferentially perpendicular to the sheet, then the
correction factors to the observed column density $N_{obs}$ and \blos\ are 
not independent.
Following the same argument as for the ensemble of magnetic field
measurements given above, we find that 
$B/N$ = (\blos/$N_{obs}$)(cos$^2 \theta$) for one line-of-sight, 
so that $B/N$ = $<3\blos/N_{obs}>$ for the average over a distribution
of equally likely line-of-sight directions.
In terms of the critical magnetic flux-to-mass ratio,
\begin{eqnarray}
\left(\frac{\Phi}{M}\right)_{\rm n} = \frac{\Phi}{M} \Bigg/
\left(\frac{\Phi}{M}\right)_{\rm cr} & = & \frac{B}{mN} \left(\frac{\sqrt{G}}{c_{\Phi}}\right)^{-1} 
\nonumber \\
& = & \digamma \, \frac{\blos}{2.8m_{\rm H}\Nhh} 
\left(\frac{\sqrt{G}}{c_{\Phi}}\right)^{-1}
\end{eqnarray}
where the subscript ``n'' indicates that the values of $\Phi/M$ have
been normalized by the critical value $(\Phi/M)_{\rm cr}$, and $\digamma$ is 
the geometric correction factor, which is 2 for
the initially uniform sphere, and 3 for the isothermal sheet.
Evaluating the constants leads to 
\begin{eqnarray}
\left(\frac{\Phi}{M}\right)_{\rm n} & = & 
2 \times 10^{20} \frac{\blos}{\Nhh} \,\, {\rm
cm^{2}\; \ug}~~{\rm (sphere)}\\
& = & 4 \times 10^{20} \frac{\blos}{\Nhh} \,\, {\rm
cm^{2}\; \ug}~~{\rm (sheet)} \,\, .
\label{eqn-shu}
\end{eqnarray}

The values of \fmr\ for the different geometries considered above are
given in Table~\ref{tbl-models}.  In this Table we also include the data for 
$B$ and $N$ compiled by Crutcher (1999) and recent Zeeman results from Sarma
\etal\ (2000; NGC~6334) and Crutcher \& Troland (2000; L1544).  
Therefore we believe Table~\ref{tbl-models} gives the most complete
summary available of Zeeman measurements in extended molecular clouds
(excluding small-scale OH maser emission).  In
column 1 is the source name, columns 2 \& 3 give the line-of-sight field
strength \blos\ and its log, and column 4 lists the log of the molecular 
hydrogen column density.  Column 5 lists the values of \fmr\
assuming an initially uniform sphere.  Column 6 lists the
values of \fmr\ assuming a sheet-like geometry.
The values in Table~\ref{tbl-models} are shown graphically in
Figure~\ref{fig-models}.  In this figure are plotted the measured field
strength \blos\ against the molecular hydrogen column density \Nhh, and
overlayed with loci of \fmr\ for the uniform sphere model (a) and the
sheet-like model (b).  

Considering the values listed in Table~\ref{tbl-models} and plotted in
Figure~\ref{fig-models}, the results for the spherical parent cloud model are 
consistent with the clouds being slightly magnetically supercritical 
(\fmr\ $<$ 1), the same result as found by Crutcher (1999), who only 
considered the detections he tabulated.  The addition of our large
number of upper limit measurements to his sample strongly reinforces
this result.
As noted by Crutcher (1999), assuming an initially uniform spherical cloud
with a uniform magnetic field may be an oversimplification.  In
particular, if clouds are supported primarily by static magnetic fields
with uniform direction they will become flattened as matter flows
inward along field lines, and the field will become pinched along the
direction perpendicular to the initial field direction.  
Therefore, the results for the sheet-like cloud may be more realistic.  
Considering only those clouds where the Zeeman effect was detected, 
Table~\ref{tbl-models} shows that the results for this geometry are consistent 
with \fmr\ $\approx$ 1, and the clouds are in approximate equilibrium between
magnetic pressure and gravity, the same result found by Shu \etal\ (1999),
though they did not distinguish between the detections and the upper
limits.  Because of the statistical corrections
applied to $B$ and $N$, and the uncertainties in measuring $N$, a more
definitive conclusion cannot be reached at present.  

If we assume for the clouds where the Zeeman effect was not detected
that the true magnetic field strengths are equal to
the 3$\sigma$ upper limit values (listed in Table~\ref{tbl-models}),
then for the sheet-like model the mean flux-to-mass ratio is $<$\fmr$>$ =
$1.3\pm0.9$ $\approx$ 1 (excluding G20.8--0.1 and G29.9+0.0 as the
sensitivity of these observations is significantly less than the other
data), the same results as for the detections.  If the 
magnetic field within the beam is uniform, the true values are not likely
to be greater than 3$\sigma$ for any of the clouds.
On Figure~\ref{fig-models} we also indicate the 1$\sigma$ limits for the
non-detections.  Assuming a uniform field, if the true field strengths are closer to the 1$\sigma$
limits than to the 3$\sigma$ limits then for the sheet-like model
the mean flux-to-mass ratio is $<$\fmr$>$ = $0.4\pm0.3$, 
significantly lower than the critical value, which implies that in general 
the clouds are magnetically supercritical by a factor 2-3.
As discussed below, field structure within the beam could reduce the
flux-to-mass ratio by a factor 3--4 over its true value.  However, beam 
dilution will also result in measurements of the column density lower than 
the true value for the gas in which the magnetic field is studied, which
will increase the flux-to-mass ratio.  

It is important to note that a number of the sources located at the upper
end of Figure~\ref{fig-models} (log \Nhh\ $>$ 22.6) are due to high
angular resolution VLA Zeeman observations.  In almost all of these cases the
initial detection of the Zeeman effect was made with low-angular
resolution single-dish 
observations.  Consequently, weaker field strengths by a factor $\sim$3--4
were reported compared
to the subsequent higher-angular resolution observations which reveal
the field structure.  Therefore there is an inconsistancy between the lower
and upper ends of Figure~\ref{fig-models}, as the lower end contains
many upper limit values from low-angular resolution observations, and
the upper end contains many clouds 
where the Zeeman effect has been detected in higher-angular resolution
inteferometric and single-dish observations.  
Observations at higher angular
resolution of sources showing only upper limits to $B$ in low-angular
resolution single-dish
observations are required.  If the lack of detection of the Zeeman
effect is due mainly to field structure within the large single-dish
beam, then these higher resolution observations should in many cases
reveal the Zeeman effect (e.g., Brogan \etal\ 1999 for M17; Sarma
\etal\ 2000 for NGC~6334) and allow us
to explore more fully the high column density range of the plot and the
structure in the field.

Table~\ref{tbl-models} also shows
that there is no clear example of a magnetically subcritical cloud, as
has been noted by Nakano (1998).  The cloud toward RCW~57 where we have
a possible detection of the Zeeman effect, and hence a measurement of $B$,
has a high value of \fmr\ for the geometries considered here, implying
that it is in a magnetically subcritical state.  As discussed earlier, high
resolution Zeeman observations of RCW~57 are required to examine the
validity of this result.

\subsubsection{Non-uniform Fields Associated with \hii\ Regions}

The foregoing results show that the flux-to-mass ratio deduced from
the observations may or may not be magnetically critical, depending on
whether the magnetic field is modeled as threading a region whose geometry
is spherical or planar.  Here we point out another geometrical effect which
can influence the conclusion of magnetic criticality.  The field direction
may not be uniform within the telescope beam, particularly for a situation
common in single-dish Zeeman observations: the molecular gas is associated with
an \hii\ region, and is seen in absorption against the \hii\ region.   In
such a case, the gas and field around the protostar may be compressed into
a thin shell by the pressure of the \hii\ region, so that the field is
largely tangent to the surface of the H II region.  The similarity of
magnetic and H II region pressures has been noted for the well-studied
source W3 OH (Reid, Myers \& Bieging 1987).  The resulting geometry differs
from the cases discussed above, since the field direction is not uniform,
but lies tangent to a layer surrounding the \hii\ region, except near the
``poles'' which mark the original field direction.  A sketch of this
situation is shown in Figure~\ref{fig-bubble}.

If the shell is perfectly spherical, a uniform field $B$ with
initial angle $\phi$ from the line-of-sight can be shown to have a
line-of-sight component, averaged over the forward hemisphere, of 
$<\blos>$ = $(2/3) B\cos \phi$; and the mean over a random distribution of 
initial angles $\phi$ is then $\ll\blos\gg$ = $B/3$.  
Thus the true field strength is a factor of 3
greater than the typical line-of-sight component. Also, this geometrical
factor should increase further if weighting due to the varying brightness
across the face of the \hii\ region is taken into account.  Then the
brightest part of the \hii\ region, at its projected center, is where the
field direction is most nearly perpendicular to the line-of-sight, and the
faintest part of the \hii\ region, at its projected edge, is where the field
lies most nearly along the line-of-sight.

If this geometrical picture is correct, one should expect to see
evidence for the corresponding field structure in higher-resolution Zeeman
maps and in submillimeter polarization maps of \hii\ regions.  The
``polarization holes'' noted in many submillimeter maps of \hii\ regions
(Hildebrand \etal\ 1995) merit study as possible counterparts of the magnetic
polar regions described here. 

In the shell model the total column density of the gas in the
forward hemisphere is unchanged from before to after the formation of the
shell, so the flux-to-mass ratio requires the same correction, by a factor
$\sim$3, as does $B$ alone.  However caution is required in applying the results
of the shell model to obtain an estimation of the flux-to-mass ratio, since the
present-day field is in the plane of the shell, rather than perpendicular
to it; and since the original flux-to-mass ratio may have changed during
the expansion of the \hii\ region, if the expansion induced some flow along
field lines.  Furthermore, as indicated by the beam-filling factors in
Table~\ref{tbl-obs}, in low-angular resolution single-dish observations
the source is generally much smaller than the beam.  This may
significantly reduce the effect of the geometry discussed here in
accounting for the low detection rate.


\section{Conclusions}

We have presented the results of a survey of 23 molecular clouds for the
Zeeman effect in OH.  For 22 of these clouds we were also able to
determine the column density.  We combined our data with the data
for 29 clouds analysed by Crutcher (1999), and compared the combined
results with simple cloud models.   

The data in Table~\ref{tbl-models} and Figure~\ref{fig-models} 
are generally consistent with a
flux-to-mass ratio less than its critical value by a factor of a few, if
the model cloud is initially a uniform sphere threaded by a field of
uniform strength and direction, independent of whether the true field
strengths for the non-detections are close to or significantly less than
the 3$\sigma$ upper limits.  
Such a magnetically supercritical cloud
should be highly unstable to collapse and fragmentation.  
These conclusions are in agreement with those of Crutcher (1999).

If instead the
model cloud is a highly flattened sheet threaded by a uniform perpendicular
field, the data are generally consistent with a critical flux-to-mass
ratio, implying a critically stable system (assuming that the true
values of $B$ for the non-detections are close to the 3$\sigma$ upper
limits).  
These conclusions are in agreement with those of Crutcher (1999) and 
those of Shu \etal\ (1999).
However, if the true values of $B$ for the non-detections are closer to
the 1$\sigma$ values, then for the sheet-like model cloud the data are
generally consistent with a flux-to-mass ratio significantly less than
critical, implying that the typical cloud is significantly magnetically
supercritical.

When both negative and positive detections of the Zeeman
effect are considered, the single-dish detection rate of the OH Zeeman
effect is relatively low, less than 10\% in the present survey.  
This low rate may be due simply to low mean field strengths, which is
the simple assumption we have used in our analysis.  
A more realistic explanation of the low detection rate may be a
selection effect which tends to decrease the field strength inferred from a
single-dish Zeeman absorption observation of the most common survey target,
an \hii\ region.  Expansion of an \hii\ region may compress the gas and field
into a shell-like configuration, decreasing the Zeeman effect compared to
that from a uniform field of the same magnitude and mean direction.
Interaction with the \hii\ region may also lead to line-of-sight reversals
of the field (\eg, Brogan \etal\ 1999), which tend to reduce the Zeeman
effect by a factor of a few when observed with coarse resolution.

These considerations imply that the flux-to-mass ratio of the
typical cloud associated with an \hii\ region may appear critical or
somewhat supercritical on the size scale to which single-dish observations
are sensitive.  But it is difficult to infer this ratio accurately from
such low-resolution observations and from models of uniform fields, because
the field is likely to have important unresolved structure.  

In summary the principal results and conclusions of this study are:

1. The Zeeman effect was detected in a molecular cloud associated with
the \hii\ region RCW~38, with a field strength of $38\pm3$ \ug.

2. The Zeeman effect was possibly detected in 
a cloud associated with the \hii\ region RCW~57, with a field
strength of $-203\pm24$ \ug.

3. If the molecular clouds are modeled as initially 
uniform density spheres with uniform magnetic fields, then generally
they are magnetically supercritical, and therefore unstable to gravitational
collapse.

4. If the molecular clouds are modeled as highly
flattened isothermal sheets, then those with detections of the Zeeman
effect are approximately magnetically critical.  
Those clouds with non-detections of the
Zeeman effect (and hence only upper limit values to the magnetic field
strength) are also approximately magnetically critical, but only if the
true field strengths are close to the 3$\sigma$ upper limits reported
here.  If instead the true field strengths are signifcantly lower, 
for example equal to the 1$\sigma$
upper limits, then on the average the clouds are approximately magnetically
supercritical by a factor 2--3.

5. For clouds associated with \hii\ regions, the molecular gas may be
swept up into a thin shell, resulting in a non-uniform magnetic field
geometry.  When this geometry is observed at low angular resolution, as
is the case with single-dish observations, field cancellation within the
beam will occur, resulting in measurements of the field strength which
are significantly lower than the true values.  

6. A number of upper limits to the field strength exist for
clouds observed in single-dish experiments.  Those observations
only sample moderate column densities and the clouds should
be observed at higher angular resolution to examine whether the lack of a
detectable Zeeman effect is due to changes in field direction within the
single-dish beam.  Such observations may hold the most promise for
further studies of magnetic field strength in local molecular clouds.

\acknowledgements

TLB thanks the SAO and the School of Physics, ADFA for financial
support.  We thank the Parkes staff for their
tremendous support, in particular Harry Fagg and Ewan Troup, and the
Green Bank staff and telescope operators, in particular Dana Balser.
We gratefully thank Robina Otrupcek for undertaking the Mopra observations.
Jonathan Williams' assistance with the Green Bank observations in 1996 is
appreciated.  
The least-squares Zeeman fitting routines were kindly provided by Carl
Heiles, and incorporated in CLASS scripts (\url{http://iram.fr/GS/gildas.html})
with help from Mario Tafalla.
We thank Tom Troland for sharing unpublished data with us.
This research has made use of NASA's Astrophysics Data System Abstract 
Service, and the SIMBAD data base, operated at CDS, Strasbourg, France.


\appendix

\section{Individual Clouds}
\label{appen-sources}

\subsection{Parkes}
\label{appen-pks}

{\sl RCW~38} -- see \S~\ref{sec-rcw38}
\skipthis{
RCW~38 is a bright \hii\ region which has recently been
observed in the near-infrared with the VLT (Alves \etal\ 2001 in prep).  In a
separate paper we present a detailed molecular line and continuum study
of the region (Bourke \etal\ 2001 in prep.) which indicates that the Zeeman
effect arises from a molecular clump on the western edge of the \hii\
region, possibly associated with the bright 10 \micron\ peak IRS1
(Smith \etal\ 1999).  Assuming \Tex\ = 10 K we determine \tauoh\ = 0.15, 
\noh\ = \power{7.6}{14}\ \cm{-2}, and \Nhh\ = \power{9.3}{21}\ \cm{-2}.  
\uco{13}\ \jj{1}{0}\ observations imply \Nhh\ = 
\power{4}{21}\ \cm{-2}\ (assuming [\hh]/[\uco{13}] = \power{7}{5}), lower 
than derived from the OH observations, which may
suggest that the OH is sampling denser gas (\eg, Roberts \etal\ 1995
for S106).  Observations of CS
\jj{2}{1}\ and \cuo{18}\ \jj{1}{0}\ support this view (Bourke \etal\ 2001
in prep.).    
}

{\sl Carina Molecular Cloud} -- Associated with $\eta$ Carina.  It has
been mapped in OH by Dickel \& Wall (1974), who
found two peaks in the absorption, closely corresponding to the dark
lanes either side of the optical emission associated with $\eta$ Carina.  The
position we have observed is the western peak.  
Brooks \etal\ (2001 in prep) are currently undertaking a multi-wavelength 
study of the 
Carina Molecular Cloud, and preliminary results from this study have been 
reported (Brooks, Whiteoak \& Storey 1998).  The distance has been determined 
by Tapia \etal\ (1988) to be 2.5 kpc.  Assuming \Tex\ = 10K we determine
\tauoh\ = 0.04 and \noh\ = \power{5.2}{14}\ \cm{-2}, which imply 
\Nhh\ = \power{6.6}{21}\ \cm{-2}, the same result we obtain from the
CO and \uco{13}\ data of Brooks \etal\ (2001 in prep).  Dickel \& Wall (1974)
find \tauoh\ = 0.033 and \noh/\Tex\ = \power{4.9}{13}\ \cm{-2}\ K$^{-1}$
at the same position.

{\sl Chamaeleon I} -- Chamaeleon I is a well studied molecular cloud
(see \eg, the review by Schwartz 1991) forming mainly low-mass stars, 
located at a distance of 160 pc (Whittet \etal\ 1997).  Though it
is easier to detect the Zeeman effect in strong absorption lines from
molecular clouds associated with \ion{H}{2}\ regions, there is a clear
need for more measurements of the magnetic field strength in cold molecular
clouds such as Chamaeleon (Crutcher \etal\ 1993), which is
why it has been included in this study.  There is evidence for a large
scale ordered magnetic field in the region (McGregor \etal\ 1994), and
this together with the bright OH lines (at least for a thermal emission source) 
makes Chamaeleon I a prime Zeeman candidate among dark clouds.  The cloud has
been mapped in OH with the Parkes radiotelescope by Toriseva \etal\ (1985) 
and the extinction has been determined throughout the entire cloud by 
Cambr\'esy \etal\ (1997) using near-infrared colours.  We assume \Tex\ =
6 K and so \tauoh\ = 0.48, \noh\ = \power{6.3}{14}\ \cm{-2}, and \Nhh\
= \power{7.8}{21}\ \cm{-2}.  For comparison, the NANTEN and DENIS data in 
the same region (Hayakawa \etal\ 1999) imply 
$N$($^{13}$CO) = \power{5-10}{15}\ \cm{-2}, or \Nhh\ =
\power{3.5-7}{21}\ \cm{-2}.  At the position we observed, Toriseva
\etal\ (1985) obtain \noh\ $\approx$ \power{7}{14}\ \cm{-2}.

{\sl RCW~57} -- see \S~\ref{sec-rcw57}
\skipthis{
Located at a distance of 3.6 kpc (kinematic, \eg, Caswell \& Haynes
1987b) or 2.4 kpc (Persi \etal\
1994, based on the possible association of HD 97499 with the molecular
cloud), RCW~57 (NGC 3576) is an \hii\ region with ongoing star formation
(\eg, Persi \etal\ 1994).  The few published studies of this region 
suggest it is a ``typical" \hii\ region with associated molecular
clouds.  The OH line profiles are presented in
Figure~\ref{fig-rcw57-zee}, and are well fitted by two Gaussian components.  
Whiteoak \& Gardner (1974), observed the 5 GHz transition of \hhco\ in 
absorption toward RCW~57 with a 4\farcm2 beam, and found two velocity 
components with almost identical 
properties to the Gaussian components we fitted to our OH data, 
suggesting that at
least part of the OH absorption arises in somewhat denser gas 
(10$^4$ cm$^{-3}$) than is usually sampled by OH (10$^3$ cm$^{-3}$).   
For the component at --22.6 \kms\ we find \tauoh\ = 0.06 and \noh\ =
\power{8.9}{14}\ \cm{-2}, or \Nhh\ = \power{1.1}{22}\ \cm{-2}\ compared
with \Nhh\ = \power{1.3}{22}\ \cm{-2}\ from \uco{13}\ Mopra observations.
This is the component where we have a probable detection of the Zeeman
effect, with \blos\ = $-203\pm24$ \ug.
For the component at --26.1 \kms\ we find \tauoh\ = 0.2 and \noh\ =
\power{1.5}{15}\ \cm{-2}, or \Nhh\ = \power{1.9}{22}\ \cm{-2}\ compared
with \Nhh\ = \power{1.5}{22}\ \cm{-2}\ from \uco{13}\ Mopra observations.
}

{\sl G326.7+0.6} --
Very little is known about this region.  The radio continuum source is probably
associated with the optical \hii\ region RCW~95.
The OH component at --44.8 \kms\ has the same velocity as the H109$\alpha$
recombination line (Caswell \& Haynes 1987b), and is associated with the
\hii\ region at a distance of 3.0 kpc (kinematic).  The narrow line
width of the --21.6 \kms\ component suggests that it arises from an 
unassociated dark cloud along the line-of-sight to the continuum source, with a
kinematic distance of 1.5 kpc.  For the --44.8 and --21.6 \kms\
components we determine \noh\ = \power{9.5}{14}\ \cm{-2}\ and
\power{3.1}{14}\ \cm{-2}\ respectively, assuming \Tex\ = 10 and 6 K,
respectively.  These values imply \Nhh\ = \power{1.2}{22}\ \cm{-2}
(compared to \Nhh\ = \power{2.3}{22}\ \cm{-2}\ from \uco{13}\
observations) and \Nhh\ = \power{3.8}{21}\ \cm{-2}\ (\Nhh\ =
\power{2}{21}\ \cm{-2}\ from \uco{13}) respectively.

{\sl G327.3--0.5} --
Possibly associated with RCW~97 at a kinematic distance of $\sim$3.3
kpc, the only detailed study of the
molecular gas associated with G327.3--0.5 has been reported by 
Brand \etal\ (1984) who mapped the cloud in the 
\uco{12}\ \jj{2}{1}\ transition, and found a massive ($\sim$10$^4$
\Msun) cloud associated with a number of infrared sources (Frogel \&
Persson 1974), and a possible bipolar molecular outflow associated with the 
southern-most group of infrared sources.  
Circularly polarized masers are present in both OH lines, and
unfortunately extend well into the blue side of the main OH
absorption line.  For the absorption line we find \noh\ =
\power{2.3}{15}\ \cm{-2}, assuming \Tex\ = 10 K, which implies \Nhh\ =
\power{2.8}{22}\ \cm{-2}, comparing favourably with that determined from
the Mopra \uco{13}\ observations of \Nhh\ = \power{3}{22}\ \cm{-2}.  
The Zeeman effect is clearly seen in three of the
masers present in the 1667 line.  However, since the properties of the
molecular gas associated with masers is poorly understood we simply
state that the magnetic field strengths we infer from the observed 
splitting (Reid \& Silverstein 1990) are +1.3, +2.9, and +7 mG, for masers 
with velocities at $-80$, $-68$ and $-52$ \kms,
respectively, though the identification of the Zeeman pair for the $-52$
\kms\ component is not certain. 

{\sl G343.4--0.0} --
Another southern \hii\ region which has not been studied, G343.4--0.0 is 
not clearly associated with any RCW region. 
The two main absorption features are at
$-27$ \kms\ (near the recombination line velocity) and the narrow, deep
line near 6 \kms.  Other weak absorption lines are seen, but the
signal-to-noise (S/N) is $<$ 50 so they are not
considered in the Zeeman analysis below.  For the same reason the $-27$
\kms\ feature is not used in the Zeeman analysis. 
We somewhat arbitrarily assign the near kinematic distance of 2.8 kpc to
the \hii\ region.  Unlike the sources discussed above, G343.4--0.0 
has not even been included in many of the major molecular line
and maser surveys of the southern galactic plane.
Whiteoak \& Gardner (1974) observed the 6 \kms\ component in
formaldehyde absorption, and OH absorption observations were reported 
in the extensive OH survey by Turner (1979).  
The 6 \kms\ cloud is
clearly local, and may be associated with the Lupus clouds, which have a
similar velocity and lie about 10\arcdeg\ above the plane at a similar
galactic longitude.  It may also be associated with the 6 \kms\
component seen toward NGC~6334 and NGC~6357 (W22; Kaz\'es \& Crutcher
1986; Crutcher \etal\ 1987; Massi, Brand \& Felli 1997) near
longitude 351\arcdeg.  Kaz\'es \& Crutcher (1986) detected the Zeeman
effect in the 6 \kms\ component toward NGC 6357 (W22B) with a magnetic
field strength of $-18 \pm 1$ \ug, while Crutcher \etal\ (1987) claim a
detection at a nearby position (W22A) with a field strength of $-32 \pm
9$ \ug.  The large scale CO maps made with the Columbia
1.2 m radiotelescope (\eg, Bronfman \etal\ 1989; Bitran \etal\ 1997),
suggest an association between all the clouds near 6 \kms.  If
associated with Lupus, a distance $\la$200 pc is appropriate.  For this
velocity component we find
\noh\ = \power{4.3}{14}\ \cm{-2}, which implies \Nhh\ = \power{5.4}{21}\
\cm{-2}\ compared with \Nhh\ = \power{2}{21}\ \cm{-2}\ derived from the
\uco{13}\ observations.

{\sl NGC 6334} --
NGC~6334 is a well studied \hii\ region located at 1.7 kpc (Neckel
1978).  Visually it is very prominent, with numerous nebulous spots, and
forms a fine pairing with the nearby NGC 6357 (W22).  Important studies
include the CO observations of Dickel, Dickel \& Wilson (1977), the OH
observations by Brooks (1995; published in Brooks \& Whiteoak 2001) and the 
recent PhD thesis by Kraemer
(1998), who studied [\ion{O}{1}], [\ion{C}{2}], CO, CS and ammonia
throughout the star-forming molecular ``ridge".  There are two main
absorption components, one at $-4$ \kms\ associated with the \hii\
region, and one at 6 \kms.  As discussed above for G343.4--0.0, the 6
\kms\ component is local and covers a large area, but has not been
studied in any detail and its distance and properties are not
well known.  Recently Sarma \etal\
(2000) presented Zeeman observations of the $-4$ \kms\ component in OH 
and \ion{H}{1}\ made with the VLA.
Following the nomenclature of Rodr\'{\i}guez, Cant\'{o} \& Moran (1982), they
measured the Zeeman effect in OH against source A (\blos = $152 \pm 17$
\ug) and in \ion{H}{1}\ against source D ($-93 \pm 13$ \ug), source E
($-180 \pm 29$ \ug), and NW of D ($169 \pm 55$ \ug).  The position we
observed covers source D and the position NW of D.  Sources A and E lie
just outside our primary beam.  The ATCA OH channel maps presented by Brooks
(1995) show strong absorption against source D, and only weak absorption
against source E and NW of D, which suggests that the absorption we
observe in the single-dish observations is mainly due to source D.  
We do not detect the Zeeman effect in OH.  It is clear that the magnetic
field changes direction within the Parkes beam, as indicated by the
signs of \blos\ for D and NW of D, so the lack of detection may be due
to field reversal and therefore cancellation within the beam.
From our observations we find \noh\ = \power{1.4}{15}\ \cm{-2}, which
implies \Nhh\ = \power{1.8}{22}\ \cm{-2}.  From our limited \uco{13}\
observations we find \Nhh\ = \power{1}{22}\ \cm{-2}, while Kraemer \& Jackson
(1999) obtain a value of \Nhh\ = \power{2.2}{22}\ \cm{-2}\ toward source D
from their CO observations.

Troland and co-workers (Troland 1995, private communication)
have searched for the Zeeman effect toward NGC 6334 with the Nan\c{c}ay
telescope and report a 3$\sigma$ upper limit for the 6 \kms\ component of 
$\sim$20 \ug.  For this cloud we find \noh\ = \power{1.2}{14}\ \cm{-2}\
and \Nhh\ = \power{2.8}{21}\ \cm{-2}, assuming \Tex\ = 6 K.  From \uco{13}\ 
observations we obtain \Nhh\ = \power{2}{21}\ \cm{-2}.

{\sl G8.1+0.2} --
The recombination line velocity for the G8.1+0.2 \hii\ region 
is $\sim$20 \kms, which implies a kinematic distance of $\sim$3.5 kpc.  
The spectra show two velocity components near 13 and 17
\kms\ respectively, and a circularly polarized maser pair 
near 22 \kms.  The S/N ratio of the 13 \kms\ component is too low to 
be considered in the Zeeman analysis.  For the 17 \kms\ component \noh\
= \power{2.9}{15}\ \cm{-2}\ and \Nhh\ = \power{3.6}{22}\ \cm{-2}.  
We have no \uco{13}\ observations for an independent check of these
values.
The Zeeman effect is clearly seen in the maser lines, from which we can 
infer a magnetic field strength in the masing region of $\sim$ +3 mG.  No
detailed study of this \hii\ region has been reported.

{\subsection{Green Bank}

{\sl W7} -- Also known as 3C 123, the absorption lines are due to the 
Taurus Molecular Cloud (Crutcher 1977).  Bregman \etal\ (1983) attempted
to detect the Zeeman effect in \ion{H}{1}\ using the WSRT, and reported
a 3$\sigma$ upper limit of 16 \ug.   From our observations we determine
\tauoh\ = 0.09 and \noh\ = \power{2.4}{14}\ \cm{-2}.
In the same region Wouterloot \& Habing (1985; their region f),
find \noh\ $\approx$ \power{8}{14}\ \cm{-2}.

{\sl NGC 2264-IRS2} -- Also known as IRAS 06382+0939, it was observed
during periods when the other sources on our list were not available.
The region has been observed in \cuo{18}\ \jj{1}{0}\ by Schwartz \etal\
(1985), who determine \Nhh\ $\approx$ \power{1}{22}\ \cm{-2}, and in 
\uco{13}\ \jj{2}{1}\ by Wilking \etal\
(1989), who find \Nhh\ = \power{2}{22}\ \cm{-2}.  From our OH
observations we obtain \Nhh\ = \power{5.3}{21}\ \cm{-2}.  IRS2 is
associated with the northern molecular core, while IRS1 (Allen's
infrared source) is found in the southern core.  Due to the weak OH
emission lines our limit to \blos\ is not as low as has been determined
for other dark molecular clouds (e.g., Crutcher \etal\ 1993).

{\sl G10.2--0.3} -- Associated with the W31 region at a distance of
$\sim$2.0 kpc.  As the Zeeman effect was not detected and we were not
able to decompose the complex line profile with the same set of
Gaussians for both OH transitions, no further analysis was performed.

{\sl G14.0--0.6} -- The OH line velocity is similar to that of the
recombination line (Lockman 1989), with a kinematic distance of $\sim$2.4 kpc.  
It may be associated with the other sources observed here with velocities
near 20 \kms, including IRAS 18153--1651 and G14.5--0.6/IRAS 18164--1645
(Jaffe, Stier \& Fazio 1982).  No studies of the molecular gas in this 
region are known.
We find \Nhh\ = \power{1.4}{22}\ \cm{-2}.  The continuum source may be
associated with IRAS 18151--1707.

{\sl IRAS 18153--1651} -- Located in the same region as G14.0--0.6, a similar
distance is assumed.  Jaffe \etal\ (1982) detected it as a far-infrared 
source (G14.21--0.53) between 40--250 \um, and from their \uco{13}\ 
observations we infer a column density of \Nhh\ = \power{5.3}{22}\ \cm{-2}.  
From our observations we find \Nhh\ = \power{1.3}{22}\ \cm{-2}.

{\sl G14.5--0.6/IRAS 18164--1645} -- Our Green Bank observations of these 
two sources overlap.  They are part of the same group of sources
located around 2.4 kpc (which includes G14.0--0.6 and IRAS 18153--1651)
and were observed by Jaffe \etal\ (1982) in \uco{13}\ (G14.5--0.6
$\equiv$ G14.33--0.64 and IRAS 18164--1645 $\equiv$ G14.43--0.69).    
Using their observations of \uco{13}\ we infer \Nhh\ = \power{6.4}{22}\
\cm{-2}\ for G14.5--0.6 and \Nhh\ = \power{5.7}{22}\ \cm{-2}\ for IRAS
18164--1645.  From our OH observations we obtain \Nhh\ =
\power{1.7}{22}\ \cm{-2}\ for G14.5--0.6 and \Nhh\ = \power{1.2}{22}\
\cm{-2} for 18164--1645.  We know of no other studies of these clouds.

{\sl G20.8--0.1} -- The \hii\ region is associated with IRAS 18264--1652,
and has a recombination line velocity of $\sim$56 \kms.  The absorption
line we see in OH is unassociated with the \hii\ region, with a velocity
of 6.8 \kms.  This velocity implies a kinematic distance of only 600 pc,
which is unreliable, but does indicate that the molecular cloud is
probably nearby.  No studies of this cloud were found in the literature.  From
our OH observations we infer \Nhh\ = \power{1.2}{21}\ \cm{-2}.

{\sl G29.9+0.0} -- The background continuum source is the very well
studied UC\hii\ region G29.96--0.02 (e.g., Pratap, Megeath \& Bergin 1999), 
with a recombination line velocity near
100 \kms.  As is the case for G20.8--0.1, our OH observations arise from
a foreground region near 8 \kms, in an anonymous molecular cloud.  Its
position and velocity suggest that it is associated with the Aquila Rift
series of clouds (Dame \etal\ 1987), which includes the well know Serpens
star-forming region at a distance of $\sim$300 pc (De Lara,
Chavarr\'{\i}a-K, \& L\'opez Molina 1991).  This may also be
true for G20.8--0.1.  From our OH observations we find \Nhh\ =
\power{2.1}{21}\ \cm{-2} along this line-of-sight.

{\sl G78.5+1.2} -- Also known as L889, OH Zeeman observations were
reported by Crutcher \etal\ (1993), who found a 3$\sigma$ upper limit of
6 \ug, much lower than our limit of 30 \ug.  We observed a position 
$\sim$17\arcmin\ away from their position.  At this position we find \Nhh\ =
\power{3.9}{21}\ \cm{-2}, while Crutcher \etal\ (1993) find \Nhh\ = 10$^{22}$
\cm{-2}\ at the position they observed.  Dickel, Seacord \& Gottesman (1977) 
observed
2 and 6 cm \hhco\ absorption toward this region and derived a maximum
column density of \Nhh\ = \power{6}{21}\ \cm{-2}.

{\sl G80.9--0.2} -- Associated with the compact source DR22, which has a
recombination line velocity near 0 \kms, implying a kinematic
distance of 3 kpc.  The OH lines are near 7 \kms, for which no molecular
line observations have been published.  From the OH observations we
determine \Nhh\ = \power{7.8}{21}\ \cm{-2}.

\section{Masers}
\label{appen-masers}

{\sl G312.1+0.3} -- (R.A., Dec.) = 14:05:06, --60:56:09 (B1950.0).  No
maser emission is evident in the OH 1665 line.  In the OH 1667 line a
single component is seen in both RCP and LCP within the absorption line.
This maser has \vlsr\ = $-46.1$ \kms\ and \dV\ = 0.42 \kms.

{\sl G327.3-0.5} -- see Appendix~\ref{appen-pks}.  The feature at 
$-80$ \kms\ has not previously been reported.

{\sl G331.4-0.0} -- (R.A., Dec.) = 16:07:14.0, --51:23:26) (B1950.0).  A
maser feature at \vlsr\ = $-62$ \kms\ with two velocity components is 
present in the OH 1665 line in both RCP and LCP (absorption is present
at \vlsr\ = $-46$ \kms).  In the OH 1667 line weak (\Ta\ $\sim$ 0.5 K) 
maser emission may be present in the absorption line in both RCP and
LCP, with \vlsr\ = $-44.6$ \kms\ and \dV\ = 0.43 \kms.


\clearpage

\begin{deluxetable}{lcccccccl}
\tabletypesize{\small}
\tablecolumns{9}
\tablewidth{0pt}
\tablecaption{Source list \label{tbl-obs}}
\tablehead{
\colhead{Name} & \colhead{R.A. (B1950)} & \colhead{Dec. (B1950)} &
\colhead{\Ta(1667)} & \colhead{\vlsr} & \colhead{$\Delta V$} &
\colhead{$fT_C$} & \colhead{$t_{int}$} & \colhead{Notes} \\
\colhead{}              & \colhead{~$h$~~$m$~~$s$}      &
\colhead{\phs\phs\arcdeg~~~\arcmin~~~\arcsec} &  \colhead{(K)} &
\colhead{(\kms)}        & \colhead{(\kms)} &
\colhead{(K)}           & \colhead{(hr)}   & \colhead{}
}
\startdata
\sidehead{Parkes}
RCW 38 & 08~57~14 & $-$47~19~42 & $-$15.0 & \phs\phn2.2 & \phn5.9 & 94 & 14.0
& (a) \\
Carina MC & 10~41~14 & $-$59~22~24 & \phn$-$2.2 & $-$25.4 & \phn5.8 & 80 & 12.5
 & \\
Cham I & 11~09~00 & $-$77~08~00 & \phs\phn1.1 & \phs\phn4.4 & \phn1.0 & -- &
16.4 & (b) \\
RCW 57 & 11~09~50 & $-$61~01~42 & \phn$-$5.8 & $-$26.1 & \phn3.4 & 45 &
\phn6.8 & \\
 & & & \phn$-$1.9 & $-$22.6 & \phn6.5 & & & \\
G326.7+0.6 & 15~41~01 & $-$53~57~54 & \phn$-$1.4 & $-$44.8 & \phn6.5 & 32 &
14.5 & \\
 & & & \phn$-$2.8 & $-$21.6 & \phn1.7 & & & \\
G327.3--0.5 & 15~49~06 & $-$54~27~06 & \phn$-$4.1 & $-$49.0 & \phn5.5 &
35 & \phn5.8 & \\
G343.4--0.0 & 16~55~43 & $-$42~31~54 & \phn$-$1.9 & \phs\phn5.7 & \phn0.6 
& \phn7 & \phn8.5 & \\
NGC6334 & 17~16~55 & $-$35~44~02 & \phn$-$6.9 & \phn$-$3.8 & \phn4.9 & 82 & 
\phn6.5 & \\
 & & & \phn$-$7.6 & \phs\phn6.3 & \phn1.2 & & & \\
G8.1+0.2 & 18~00~00 & $-$21~48~12 & \phn$-$1.0 & \phs17.4 & \phn2.8 & 12 & 
\phn4.3 & \\
\sidehead{Green Bank}
W7 & 04~33~45 & \phs29~32~00 & \phn$-$0.6 & \phs\phn4.7 & \phn2.0 & 10 
& 31.2 & \\
NGC2264--IRS2 & 06~38~12 & \phs09~39~00 & \phn\phs0.3 & \phs\phn5.4 
& \phn2.7 & \phn2 & 24.0 & (b,c) \\
G10.2--0.3 (W31) & 18~06~20 & $-$20~19~10 & \phn$-$2.1 & \phs11.4 & 11.8
& 29 & 17.2 & (a) \\
G14.0--0.6 & 18~15~10 & $-$17~06~00 & \phn$-$0.8 & \phs20.7 & \phn5.5 
& 13 & 21.9 & \\
IRAS 18153--1651 & 18~15~18 & $-$16:51:00 & \phn$-$1.1 & \phs20.1 &
\phn4.3 & 16 & 11.0 & \\
G14.5--0.6 & 18~16~06 & $-$16~50~34 & \phn$-$1.2 & \phs20.2 & \phn3.7 &
12 & \phn2.8 & (d) \\
IRAS 18164--1645 & 18~16~24 & $-$16~45~00 & \phn$-$1.0 & \phs20.1 &
\phn3.4 & 12 & \phn9.8 & (d) \\
G20.8--0.1 & 18~26~30 & $-$10~53~43 & \phn$-$0.5 & \phs\phn6.8 & \phn1.4
& 15 & \phn6.4 & \\
G29.9+0.0 & 18~43~28 & $-$02~44~02 & \phn$-$0.6 & \phn\phs7.6 & \phn3.1
& 23 & 10.0 & \\
G78.5+1.2 & 20~24~14 & \phs40~02~03 & \phn$-$1.4 & \phs\phn0.5 & \phn2.0
& 18 & 10.8 & (e) \\
G80.9--0.2 & 20~38~00 & \phs41~05~10 & \phn$-$0.9 & \phn\phs6.4 & \phn3.2 &
10 & 27.0 & \\
\enddata
\tablenotetext{(\rm a)}{Not well represented by a Gaussian}
\tablenotetext{(\rm b)}{Emission line}
\tablenotetext{(\rm c)}{IRAS 06382+0939}
\tablenotetext{(\rm d)}{Beams overlaps}
\tablenotetext{(\rm e)}{L889, nearby position observed by Crutcher \etal\ 1993}
\vspace*{-12pt}
\tablecomments{For sources where the electron temperature ($T_e$) has
been measured from radio recombination lines observations, the source
filling factor $f$ can be estimated.  The average $T_e$ 
is $\sim$6500~K, which implies a range in $f$ of 0.002-0.013.
Similar results are found for sources where a size estimate is
available.}
\end{deluxetable}

\clearpage

\begin{deluxetable}{lcccccl}
\tablewidth{0pt}
\tablecolumns{7}
\tablecaption{Magnetic Field Strengths 
\label{tbl-zee1}}
\tablehead{
\colhead{Name} & \colhead{\vlsr} & 
\colhead{\phn\phn$B_{1665}$}  & \colhead{\phn\phn$B_{1667}$} &
\colhead{\phn\phn\blos\ $\pm$ $\sigma_B$} & \colhead{\hii} & 
\colhead{Notes}  \\
 & \colhead{(\kms)} & \colhead{\phn\phn(\ug)} & \colhead{\phn\phn(\ug)}
 & \colhead{\phn\phn(\ug)} & & 
}
\startdata
{\bf RCW 38} & \phn\phs1.4 & \phs\phn$38\pm3$\phn & \phs\phn$37\pm5$\phn 
& \phs\phn${\bf 38\pm3}$\phn & y & (a)\\
Carina MC & $-$25.4 & \phn\phn$-4\pm12$ & \phn\phn\phs$9\pm21$ &
\phn\phn$-1\pm10$ & y & \\
Cham I & \phn\phs4.4 & \phs\phn\phn$3\pm5$\phn & \phs\phn\phn$3\pm5$\phn &
\phs\phn\phn$3\pm4$\phn & & \\
RCW 57 & $-$26.1 & \phn$-17\pm5$\phn & \phn\phn$-2\pm9$\phn &
\phn$-13\pm4$\phn & y & (b) \\
{\bf RCW 57} & $-$22.6 & $-198\pm32$ & $-210\pm38$ & ${\bf -203\pm24}$ & y &
(b) \\
G326.7+0.6 & $-$44.8 & \phn$-22\pm14$ & \phn$-32\pm20$ & \phn$-25\pm11$
& y & \\
 & $-$21.6 & \phn\phn$-6\pm6$\phn & \phn\phn\phs$5\pm6$\phn &
\phn\phn$-1\pm4$\phn & & \\
G327.3--0.5 & $-$49.0 & \phn($-21\pm7$)\phn & \phn$-11\pm4$\phn 
& \phn$-13\pm4$\phn & y & (c) \\
G343.4--0.0 & \phn\phs5.7 & \phn\phn$-1\pm5$\phn & \phn\phn$-5\pm4$\phn
& \phn\phn$-3\pm3$\phn & & \\
NGC 6334 & \phn$-$3.8 & \phn\phs$17\pm6$\phn & \phn\phs$13\pm6$\phn &
\phn\phs$15\pm4$\phn & y & \\
 & \phn\phs6.3 & \phn\phn$-2\pm6$\phn & \phn$-10\pm6$\phn &
 \phn\phn$-6\pm4$\phn & & \\
G8.1+0.2 & \phs17.4 & \phn\phs$35\pm23$ & \phn\phs$33\pm24$ &
\phn\phs$34\pm17$ & y & \\
W7 & \phn\phs4.3 & \phn\phs$10\pm17$ & \phn$-15\pm11$ &
\phn\phn$-8\pm9$\phn & & (d)\\
NGC2264--IRS2 & \phn\phs5.4 & \phn\phs$25\pm19$ & \phn\phs$13\pm27$ &
\phn\phs$21\pm16$ & & \\
G10.2--0.3 & \phs11.4 & \phn\phs$40\pm15$ & \phn\phn\phs$5\pm18$ &
\phn\phs$26\pm12$ & y & (e)\\
G14.0--0.6 & \phs20.7 & \phn\phn$-1\pm26$ & \phs$135\pm27$ 
& \phn\phs$64\pm18$ & y & \\
IRAS 18153--1651 & \phs20.1 & \phn\phn$-1\pm26$ & \phn$-43\pm23$ &
\phn$-25\pm17$ & y & \\
G14.5--0.6 & \phs20.2 & \phs\phn\phn$8\pm37$ & \phn$-32\pm20$ &
\phn$-23\pm18$ & y & \\
IRAS 18164--1645 & \phs20.1 & \phn$-16\pm25$ & \phn\phs$26\pm22$ &
\phn\phn\phs$8\pm17$ & y & \\
G20.8--0.1 & \phn\phs6.8 & \phs\phn$59\pm37$ & \phs\phn$24\pm30$ &
\phn\phn$-9\pm23$ & & \\
G29.9+0.0 & \phn\phs7.6 & \phn\phs$39\pm34$ & \phn\phs$19\pm29$ &
\phs\phn$30\pm26$ & & \\
G78.5+1.2 & \phs\phn0.5 & \phn$-27\pm16$ & \phs\phn$19\pm12$ &
\phs\phn\phn$2\pm10$ & & \\
G80.9--0.2 & \phs\phn6.4 & \phs\phn$21\pm16$ & \phs\phn$20\pm11$ &
\phs\phn$20\pm9$\phn & & \\
\enddata
\tablenotetext{(\rm a)}{Method of CK83 used to infer properties of
Gaussian subcomponent from integrated $V$ spectrum.  The line width of
this Gaussian is \dV\ = 2.3 \kms}
\tablenotetext{(\rm b)}{Method of Heiles (1988) and Goodman \& Heiles
(1994) used, as the Gaussian components overlap.  The Zeeman
fitting is performed simultaneously on both velocity components.}
\tablenotetext{(\rm c)}{The 1665 MHz result is listed in parentheses, as
interference prevents a meaningful fit.}
\tablenotetext{(\rm d)}{Two velocity components.  Only one component is
fitted, as the other is too narrow.}
\tablenotetext{(\rm e)}{Complex line not well represented by Gaussian
components.  Results reported here are for the whole line profile.}
\end{deluxetable}

\clearpage

\begin{deluxetable}{lcccccl}
\tablewidth{0pt}
\tablecolumns{7}
\tablecaption{Molecular Cloud Analysis
\label{tbl-models}}
\tablehead{
\colhead{Name} & 
\colhead{\blos} & \colhead{Log $B$} & \colhead{Log $N$} &
\colhead{\fmr} & \colhead{\fmr} & \colhead{Notes} \\
\colhead{} & \colhead{(\ug)} & \colhead{(\ug)} & \colhead{(\Nhh\ \cm{-1})} &
\colhead{(Sphere)} & \colhead{(Sheet)} & 
\colhead{}
}
\startdata
\sidehead{Parkes}
RCW 38 & \phn\phn\phm{$<$} 38 & \phm{$<$} 1.58 & 21.97 & \phn\phm{$>$} 0.8 
& \phn\phm{$>$} 1.6 & \phm{(a)} \\
RCW 57 (v2) & \phn\phm{$<$} 203 & \phm{$<$} 2.31 & 22.04 & \phn\phm{$>$} 3.7 
& \phn\phm{$>$} 7.3 & (a) \\
Carina MC & \phn\phn$<30$ & $<1.48$ & 21.82 & \phn$<0.9$ & \phn$<1.8$ & \\
Cham I & \phn\phn$<12$ & $<1.08$ & 21.89 & \phn$<0.3$ & \phn$<0.6$ & \\
RCW 57 (v1) & \phn\phn$<12$ & $<1.08$ & 22.29 & \phn$<0.1$ & \phn$<0.2$ & (a) \\
G326.7+0.6 (v1) & \phn\phn$<42$ & $<1.62$ & 22.08 & \phn$<0.7$ & \phn$<1.4$ & (b) \\
G326.7+0.6 (v2) & \phn\phn$<13$ & $<1.11$ & 21.58 & \phn$<0.7$ & \phn$<1.4$ & (b) \\
G327.3--0.5 & \phn\phn$<30$ & $<1.48$ & 22.45 & \phn$<0.2$ & \phn$<0.4$ & \\
G343.4--0.0 & \phn\phn\phn$<9$ & $<0.95$ & 21.73 & \phn$<0.3$ & \phn$<0.7$ & \\
NGC 6334 (v1) & \phn\phn$<12$ & $<1.08$ & 22.25 & \phn$<0.1$ & \phn$<0.3$ & (c) \\
NGC 6334 (v2) & \phn\phn$<12$ & $<1.08$ & 21.46 & \phn$<0.8$ & \phn$<1.6$ & (c) \\
G8.1+0.2 & \phn\phn$<36$ & $<1.56$ & 22.56 & \phn$<0.2$ & \phn$<0.4$ & \\
\sidehead{Green Bank}
W7 & \phn\phn$<18$ & $<1.26$ & 21.50 & \phn$<1.1$ & \phn$<2.3$ & \\
N2264--IRS2 & \phn\phn$<48$ & $<1.68$ & 21.73 & \phn$<1.8$ & \phn$<3.5$ & \\
G14.0--0.6 & \phn\phn$<56$ & $<1.75$ & 22.15 & \phn$<0.8$ & \phn$<1.6$ & \\
IRAS 18153--1651 & \phn\phn$<51$ & $<1.71$ & 22.10 & \phn$<0.8$ & \phn$<1.6$ & \\
G14.5--0.6 & \phn\phn$<54$ & $<1.73$ & 22.22 & \phn$<0.7$ & \phn$<1.3$ & \\
IRAS 18164--1645 & \phn\phn$<51$ & $<1.71$ & 22.08 & \phn$<0.9$ & \phn$<1.7$ & \\
G20.8--0.1 & \phn\phn$<69$ & $<1.84$ & 21.09 & $<11.2$ & $<22.1$ & \\
G29.9+0.0 & \phn\phn$<78$ & $<1.89$ & 21.33 & \phn$<7.0$ & $<14.0$ & \\
G78.5+1.2 & \phn\phn$<30$ & $<1.48$ & 21.59 & \phn$<1.5$ & \phn$<3.0$ & \\
G80.9--0.2 & \phn\phn$<27$ & $<1.43$ & 21.89 & \phn$<0.7$ & \phn$<1.4$ & \\
\sidehead{Crutcher 1999}
W3 OH & \phm{$<$} 3100 & \phm{$<$} 3.49 & 23.7\phn & \phn\phm{$>$} 1.2 
& \phn\phm{$>$} 2.4 & \\
DR 21 OH1 & \phn\phm{$<$} 710 & \phm{$<$} 2.85 & 23.6\phn & \phn\phm{$>$} 0.4 
& \phn\phm{$>$} 0.7  & \\
Sgr B2 & \phn\phm{$<$} 480 & \phm{$<$} 2.68 & 23.4\phn & \phn\phm{$>$} 0.4 
& \phn\phm{$>$} 0.8 & \\
M17 SW & \phn\phm{$<$} 450 & \phm{$<$} 2.65 & 23.1\phn & \phn\phm{$>$} 0.7 
& \phn\phm{$>$} 1.4 & \\
W3 (main) & \phn\phm{$<$} 400 & \phm{$<$} 2.60 & 23.2\phn & \phn\phm{$>$} 0.5 
& \phn\phm{$>$} 1.0 & \\
S106 & \phn\phm{$<$} 400 & \phm{$<$} 2.60 & 22.8\phn & \phn\phm{$>$} 1.3 
& \phn\phm{$>$} 2.5 & \\
DR 21 OH2 & \phn\phm{$<$} 360 & \phm{$<$} 2.56 & 23.3\phn & \phn\phm{$>$} 0.4 
& \phn\phm{$>$} 0.7 & \\
OMC-1 & \phn\phm{$<$} 360 & \phm{$<$} 2.56 & 23.2\phn & \phn\phm{$>$} 0.5 
& \phn\phm{$>$} 0.9 & \\
NGC 2024 & \phn\phn\phm{$<$} 87 & \phm{$<$} 1.9\phn & 22.9\phn 
& \phn\phm{$>$} 0.2 & \phn\phm{$>$} 0.4 & \\
S88 B & \phn\phn\phm{$<$} 69 & \phm{$<$} 1.84 & 22.3\phn & \phn\phm{$>$} 0.7 
& \phn\phm{$>$} 1.4 & \\
B1 & \phn\phn\phm{$<$} 27 & \phm{$<$} 1.43 & 21.9\phn & \phn\phm{$>$} 0.7 
& \phn\phm{$>$} 1.3 & \\
W49 B & \phn\phn\phm{$<$} 21 & \phm{$<$} 1.32 & 21.6\phn & \phn\phm{$>$} 1.1 
& \phn\phm{$>$} 2.1 & \\
W22 & \phn\phn\phm{$<$} 18 & \phm{$<$} 1.26 & 22.2\phn & \phn\phm{$>$} 0.2  
& \phn\phm{$>$} 0.5 & \\
W40 & \phn\phn\phm{$<$} 14 & \phm{$<$} 1.15 & 22.\phn\phn & \phn\phm{$>$} 0.3 
& \phn\phm{$>$} 0.6 & \\
$\rho$ Oph 1 & \phn\phn\phm{$<$} 10 & \phm{$<$} 1.\phn\phn & 21.7\phn 
& \phn\phm{$>$} 0.4 & \phn\phm{$>$} 0.8 & \\
L1544 & \phn\phn\phm{$<$} 11 & \phm{$<$} 1.04 & 22.3\phn & \phn\phm{$>$} 0.1 
& \phn\phm{$>$} 0.2 & (d) \\
NGC 6334 & \phn\phm{$<$} 150 & \phm{$<$} 2.18 & 22.7\phn & \phn\phm{$>$} 0.6 
& \phn\phm{$>$} 1.2 & (e) \\
OMC-N4 & \phn$<300$ & $<2.48$ & 23.1\phn & \phn$<0.5$ & \phn$<0.9$ & (f) \\
Tau G & \phn\phn\phn$<7$ & $<0.85$ & 21.6\phn & \phn$<0.4$ & \phn$<0.7$ & (f) \\
L183 & \phn\phn$<15$ & $<1.18$ & 21.2\phn & \phn$<1.9$ & \phn$<3.7$ & (f) \\
L1647 & \phn\phn$<11$ & $<1.04$ & 22.1\phn & \phn$<0.2$ & \phn$<0.3$ & (f) \\
$\rho$ Oph 2 & \phn\phn$<13$ & $<1.11$ & 21.6\phn & \phn$<0.7$ & \phn$<1.3$ & (f) \\
TMC-1 & \phn\phn$<12$ & $<1.08$ & 21.9\phn & \phn$<0.3$ & \phn$<0.6$ & (f) \\
L1495W & \phn\phn\phn$<9$ & $<0.95$ & 21.6\phn & \phn$<0.5$ & \phn$<0.9$ & (f) \\
L134 & \phn\phn$<10$ & $<1.\phn\phn$ & 21.3\phn & \phn$<1.0$ & \phn$<2.0$ & (f) \\
TMC-1C & \phn\phn\phn$<9$ & $<0.95$ & 21.9\phn & \phn$<0.2$ & \phn$<0.5$ & (f) \\
L1521 & \phn\phn\phn$<9$ & $<0.95$ & 21.7\phn & \phn$<0.4$ & \phn$<0.7$ & (f) \\
L889 & \phn\phn\phn$<6$ & $<0.78$ & 22.\phn\phn & \phn$<0.1$ & \phn$<0.2$ & (f) \\
Tau 16 & \phn\phn\phn$<7$ & $<0.85$ & 21.7\phn & \phn$<0.3$ & \phn$<0.6$ & (f) \\
\enddata
\tablenotetext{(\rm a)}{(v1) = --26.1 \kms\ component, (v2) = --22.6 \kms.}
\tablenotetext{(\rm b)}{(v1) = --44.8 \kms\ component, (v2) = --21.6 \kms.}
\tablenotetext{(\rm c)}{(v1) = --3.8 \kms\ component, (v2) = 6.3 \kms.}
\tablenotetext{(\rm d)}{From Crutcher \& Troland(2000)}
\tablenotetext{(\rm e)}{From Sarma \etal\ (2000).  Corresponds to our
NGC 6334 (v1).}
\tablenotetext{(\rm f)}{Note that the values listed
here for \blos\ are 3$\sigma_B$ for consistency with our other results,
rather than \blos\ + 3$\sigma$ as in Crutcher (1999).}
\end{deluxetable}

\clearpage 


\figcaption[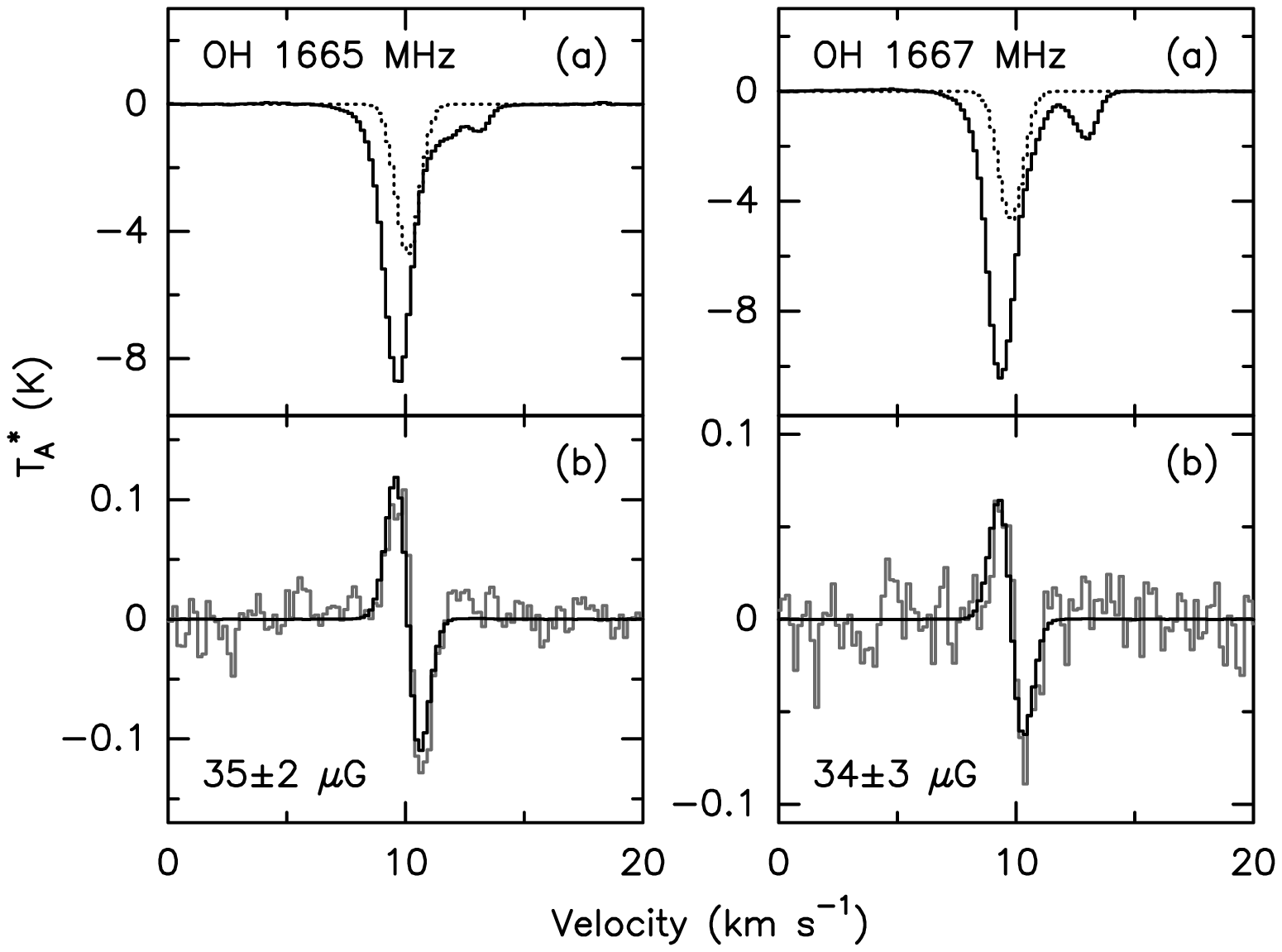]
{The Zeeman effect toward Orion~B, as observed with the Parkes radio
telescope in October 1996.
On the left is the data for the OH 1665 line, on the right the OH 1667 line.  
In (a) is the OH line profile
(average of right- and left-circular polarizations; solid line) with 
the Gaussian component responsible for the Zeeman
effect indicated (dotted line).  In (b) is the Stokes $V$
spectrum (right- minus left-circular polarization; light grey line), with the
derivative of the Gaussian fitted to derive the magnetic field strength
(continuous black line).  The value of \blos\ inferred from the fits are 
indicated.
\label{fig-orionb} }

\figcaption[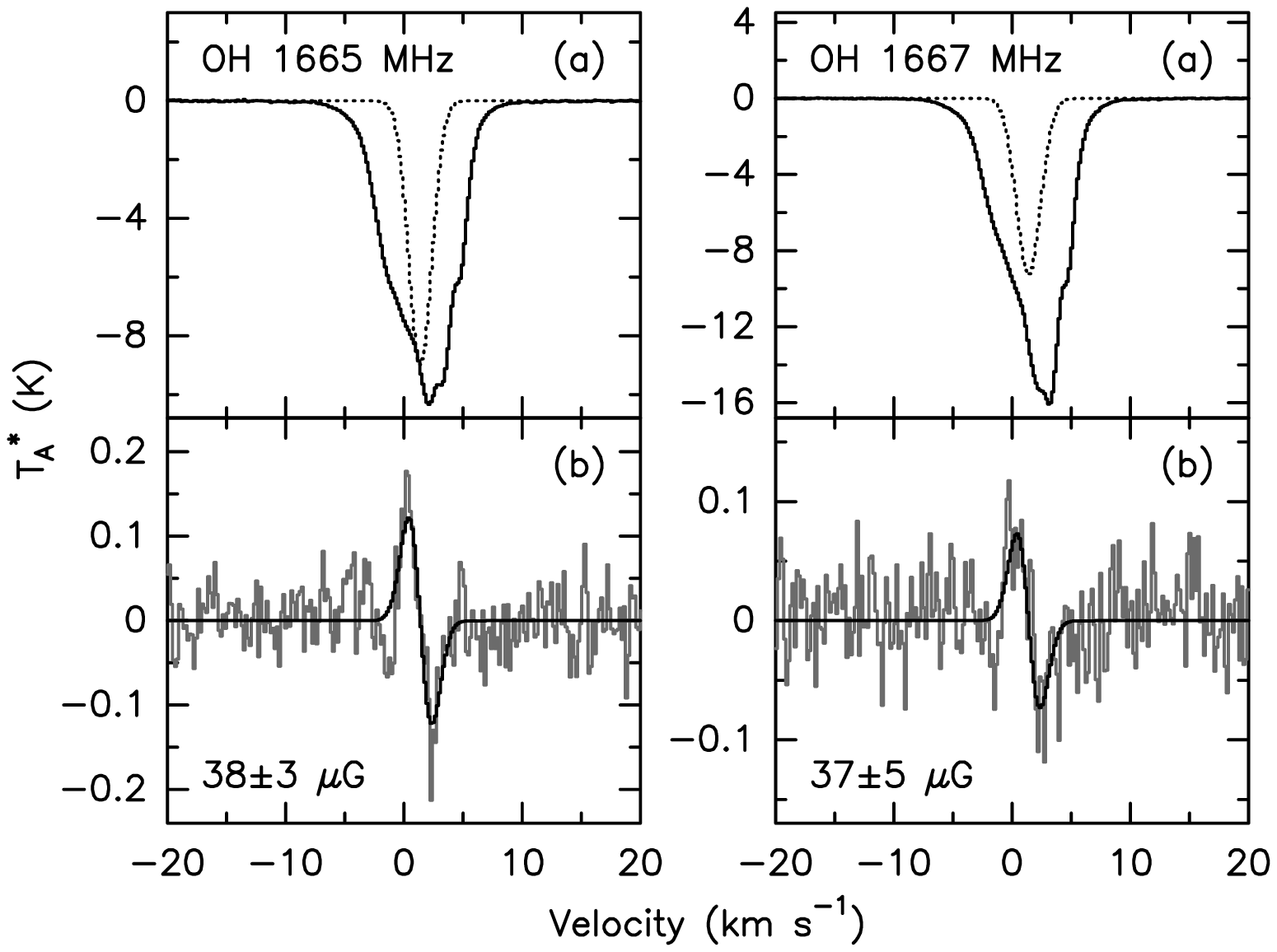]
{The Zeeman effect toward RCW~38, as observed in October 1996.
On the left is the data for the OH 1665 line, on the right the OH 1667 line.  
In (a) is the OH line profile
(average of right- and left-circular polarizations; solid line) with 
the Gaussian component responsible for the Zeeman
effect indicated (dotted line).  In (b) is the Stokes $V$
spectrum (right- minus left-circular polarization; light grey line), with the
derivative of the Gaussian fitted to derive the magnetic field strength
(continuous black line).  The value of \blos\ inferred from the fits are 
indicated.
The OH 1665 and 1667 Gaussians are assumed to have the same 
line depth, and so are optically thick.
\label{fig-rcw38}}

\figcaption[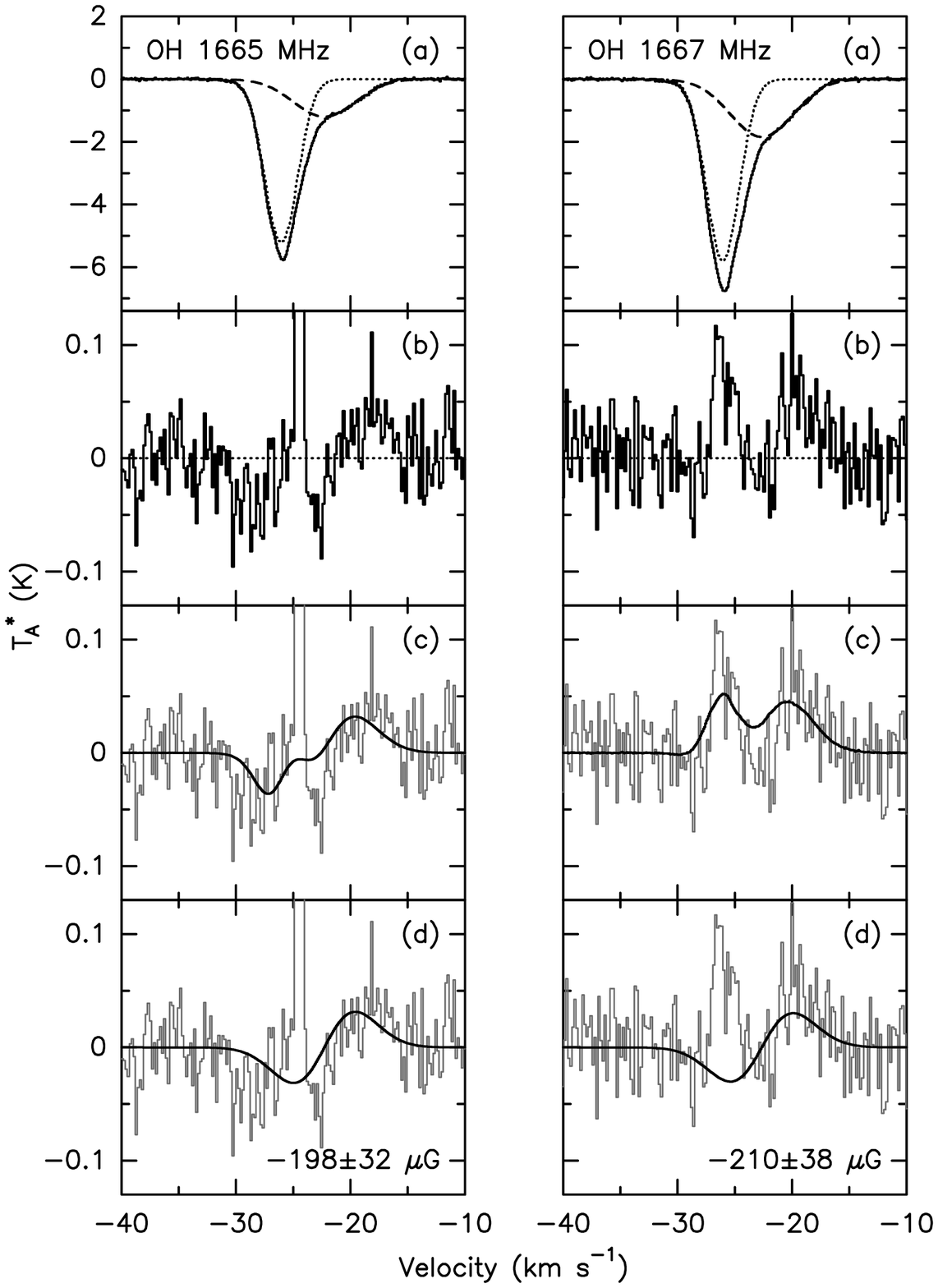]{The Zeeman effect toward RCW~57.
On the left is the data for the OH 1665 line, on the right the OH 1667 line.  
In (a) is  the OH line profile
(average of right- and left-circular polarizations; solid line) with 
the Gaussian components whose sum best fits the line profile indicated 
(dotted and dashed lines).  In (b) is the Stokes $V$
spectrum (right- minus left-circular polarization).  Maser emission is
present in the OH 1665 $V$ spectrum at velocities near $-25$ \kms.  The
feature near $-27$ \kms\ in the OH 1667 spectrum is probably due to gain
differences between the right- and left-circular polarizations, though
very weak maser emission cannot be ruled out.  
In (c) is the $V$ spectrum (light spectrum) with 
the best overall fit (dark line), consisting of the sum of the derivatives 
of the Gaussian components
shown in (a), and a gain term, which is a scaled version of the OH line 
profile.  The maser emission in the OH 1665 $V$ spectrum has been
excluded when fitting the spectrum.   The feature near $-27$ \kms\ in
the OH 1667 spectrum appears to be well fitted by the gain term
component in the overall fit.
In (d) is shown the $V$ spectrum (light spectrum) overlayed with the
derivative of the Gaussian component at $-22.6$ \kms, which is one
component of the fits shown in (c).  The value of the \blos\ inferred
from this component is indicated at the bottom of the panel.
\label{fig-rcw57-zee}}

\figcaption[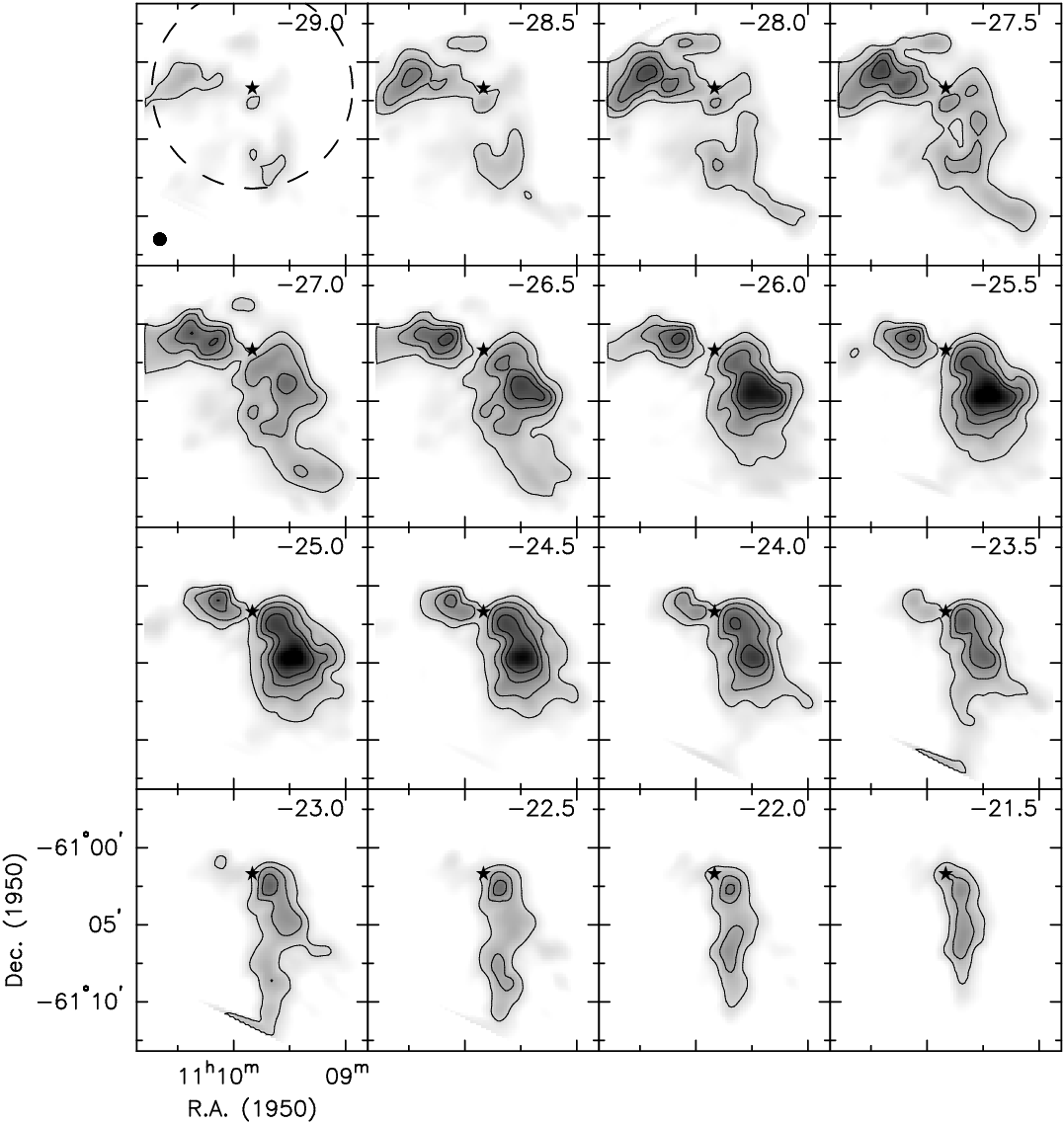]
{\uco{13}\ \jj{1}{0}\ channel maps for RCW~57.  The central velocity of
each panel is indicated in the upper right.  The Mopra beam size is
shown in the lower left corner of the upper left panel as a filled circle, 
and the Parkes OH beam size
is indicated by the dashed circle, which is centered on the position
observed for the Zeeman effect (indicated with a star).  The greyscale
ranges from 1 to 12 K \kms\ in \Ta.  The contour levels are 30, 50, 70,
and 90\% of 12 K \kms.
\label{fig-rcw57-13co}}

\figcaption[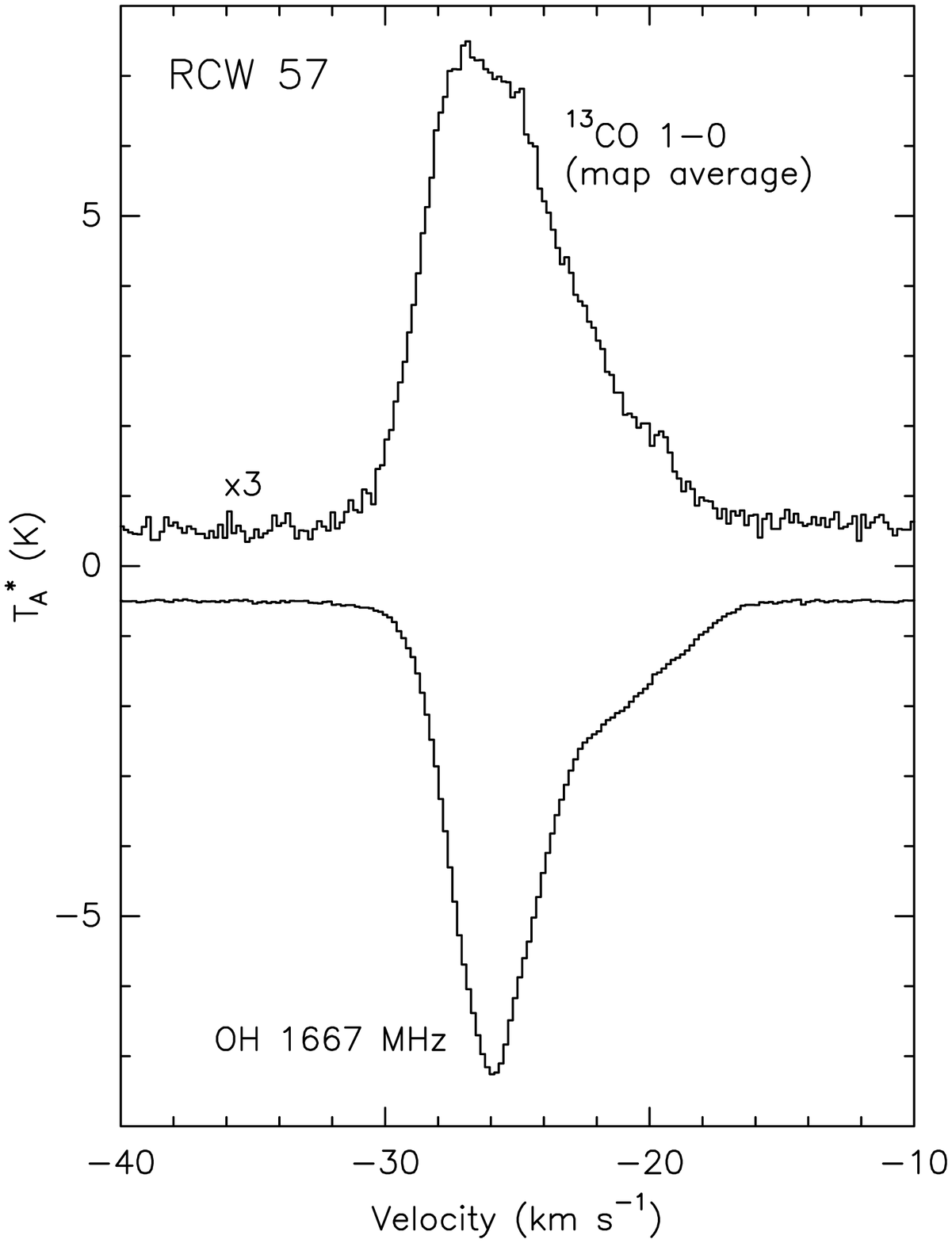]
{Average of RCW~57 \uco{13}\ \jj{1}{0}\ spectra compared with the OH
1667 MHz absorption spectrum.  The \uco{13}\ spectrum is offset by 0.5 K
and the OH spectrum by $-0.5$ K.  The map of \uco{13}\ \jj{1}{0}\
emission is shown in Figure~\ref{fig-rcw57-13co}, with the area covered
by the OH observations.
\label{fig-rcw57-13co-oh}}

\figcaption[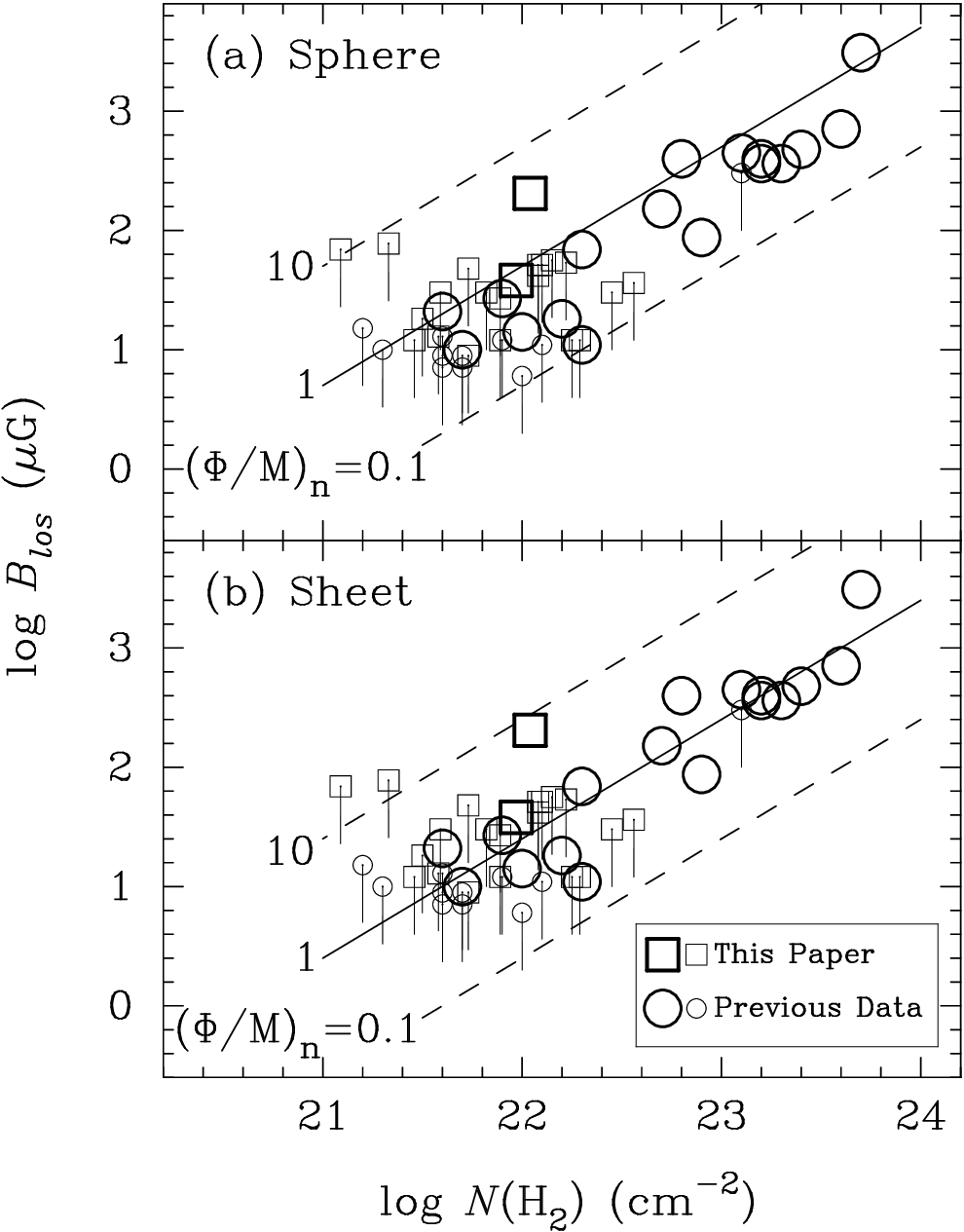]{Magnetic field strength \blos\ plotted against column
density \Nhh\ for the sources observed in this paper (squares) and for
previously published results (circles; Crutcher 1999; Sarma \etal\ 2000;
Crutcher \& Troland 2000).  The large symbols represent detections of
the Zeeman effect, while the small symbols represent 3$\sigma$ upper
limits.  The lines drawn down from the upper limits represent the shift
of the 3$\sigma$ upper limits to the 1$\sigma$ limits for each source.
In (a) are shown the loci for different values of the
magnetic flux-to-mass ratio, assuming an initial configuration of a
uniform density sphere, as described in Crutcher
(1999).  The values have been normalized to the critical value, so that
values $>$ 1 are subcritical, and $<$ 1 are supercritical.
In (b) are shown loci assuming an infinite sheet geometry
(``highly flattened molecular cloud''), like that discussed in Shu \etal\
(1999).
\label{fig-models}}

\figcaption[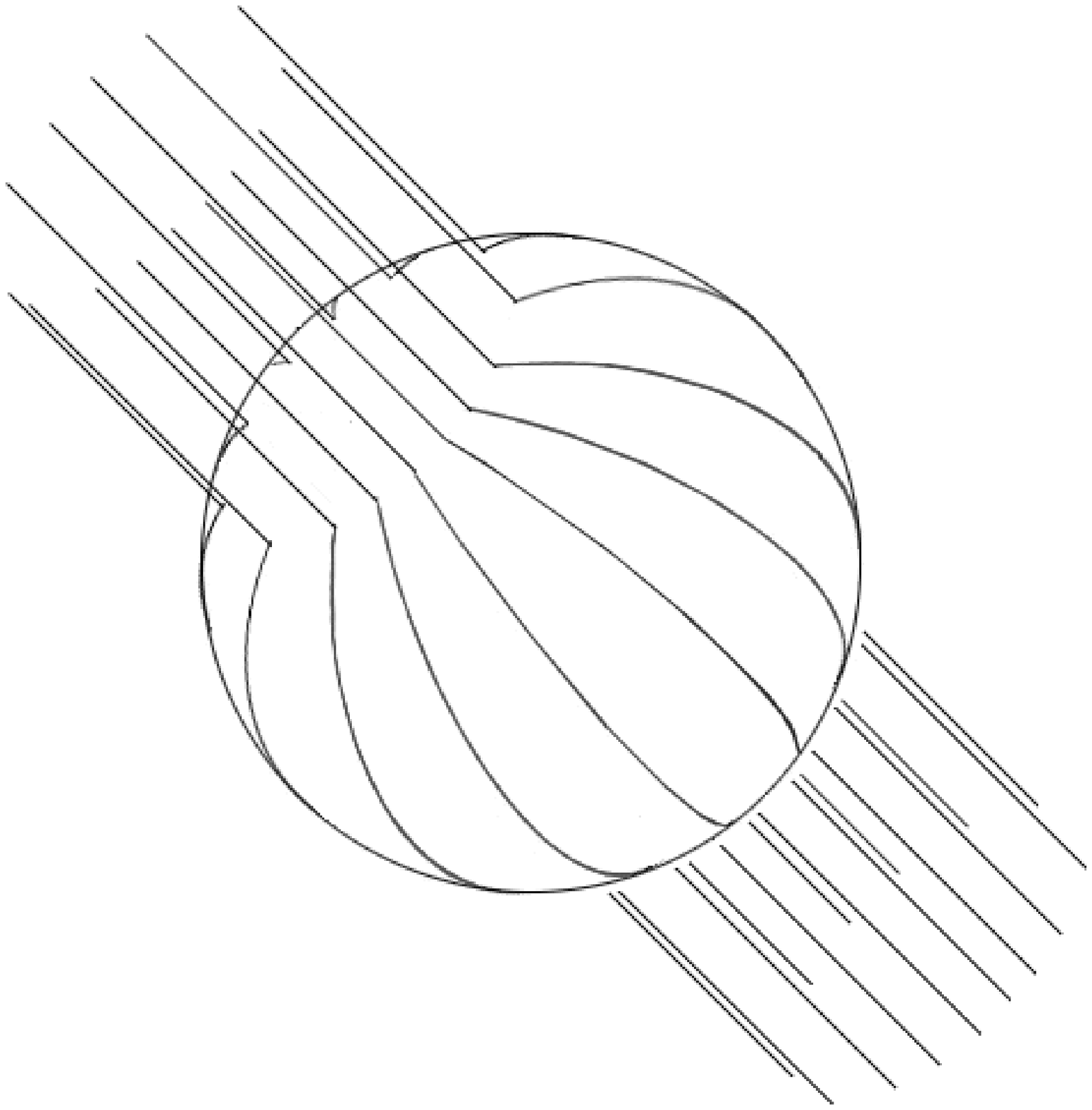]
{Sketch of magnetic field configuration near a spherical \hii\
region.  The field lines are initially uniform, and retain their original
distribution far from the \hii\ region.  As the \hii\ region develops, it
expands and compresses the surrounding molecular gas and magnetic field
into a thin spherical shell.  This configuration implies a low detection
rate for Zeeman absorption observations made with a beam larger than the
\hii\ region.  For any line-of-sight direction the line-of-sight field 
component, averaged over
the forward hemisphere, is substantially less than the line-of-sight
component of a uniform field having the same mean magnitude and direction.
\label{fig-bubble}}


\begin{figure}
\figurenum{\ref{fig-orionb}}
\epsscale{1.0}
\plotone{f1.eps}
\caption{}
\end{figure}

\begin{figure}
\figurenum{\ref{fig-rcw38}}
\epsscale{1.0}
\plotone{f2.eps}
\caption{}
\end{figure}

\begin{figure}
\figurenum{\ref{fig-rcw57-zee}}
\epsscale{1.0}
\plotone{f3.eps}
\caption{}
\end{figure}

\begin{figure}
\figurenum{\ref{fig-rcw57-13co}}
\epsscale{1.0}
\plotone{f4.eps}
\caption{}
\end{figure}

\begin{figure}
\figurenum{\ref{fig-rcw57-13co-oh}}
\epsscale{1.0}
\plotone{f5.eps}
\caption{}
\end{figure}

\begin{figure}
\figurenum{\ref{fig-models}}
\epsscale{1.0}
\plotone{f6.eps}
\caption{}
\end{figure}

\begin{figure}
\figurenum{\ref{fig-bubble}}
\epsscale{1.0}
\plotone{f7.eps}
\caption{}
\end{figure}


\begin{references}
\reference{}{Allen, A., \& Shu, F. H. 2000, \apj, 536, 368}
\reference{}{Alves, J., Petr, M., Muench, A. A., Lada, C. J., Lada, E. A., 
Moorwood, Cuby, J. G.  2001, in prep}
\reference{}{Bitran, M., Alvarez, H., Bronfman, L., May, J., \&
Thaddeus, P. 1997, \aaps, 125, 99}
\reference{}{Bourke, T. L., Myers, P. C., Robinson, G., \& Hyland, A.
R. 2001, in prep}
\reference{}{Brand, J., van der Bij, M. D. P., de Vries, C. P., Israel,
F. P., de Graauw, T., van de Stadt, H., Wouterloot, J. G. A., Leene, A.,
\& Habing, H. J. 1984, \aap, 139, 181}
\reference{}{Bregman, J. D., Forster, J. R., Troland, T. H., Schwartz,
U. J., Goss, W. M., \& Heiles, C. 1983, \aap, 118, 157}
\reference{}{Brogan, C. L., Troland, T. H., Roberts, D. A., \&
Crutcher, R. M. 1999, \apj, 515, 314}
\reference{}{Bronfman, L., Alvarez, H., Cohen, R. S., \& Thaddeus, P.
1989, \apjs, 71, 481}
\reference{}{Brooks, K. J. 1995, Honours Thesis, Univ. Wollongong}
\reference{}{Brooks, K. J., \& Whiteoak, J. B. 2001, \mnras, 320, 465}
\reference{}{Brooks, K. J., Whiteoak, J. B., \& Storey, J. W. V. 1998,
Publ. Astron. Soc. Aust., 15, 202}
\reference{}{Brooks, K. J., Whiteoak, J. B., \& Storey, J. W. V. 2001,
in prep}
\reference{}{Cambr\'esy, L., Epchtein, N., Copet, E., de Batz, B.,
Kimeswenger, S., Le Bertre, T., Rouan, D., \& Tiph\`ene, D. 1997, \aap,
324, L5}
\reference{}{Carpenter, J. M., Snell, R. L., Schloerb, F. P. 1990, \apj,
362, 147}
\reference{}{Caswell, J. L., \& Haynes, R. F. 1983, Aust. J. Phys., 36, 361}
\reference{}{Caswell, J. L., \& Haynes, R. F. 1987a, Aust. J. Phys., 40, 215}
\reference{}{Caswell, J. L., \& Haynes, R. F. 1987b, \aap, 171, 261}
\reference{}{Caswell, J. L., Haynes, R. F., \& Goss, W. M. 1980, 
Aust. J. Phys., 33, 639}
\reference{}{Caswell, J. L., \& Robinson, B. J. 1974, Aust. J. Phys., 27, 597}
\reference{}{Chan, S. J., Henning, Th., \& Schreyer, K. 1996, \aaps,
115, 285}
\reference{}{Crutcher, R. M. 1977, \apj, 216, 308}
\reference{}{Crutcher, R. M. 1979, \apj, 234, 881}
\reference{}{Crutcher, R. M. 1994, in ASP Conf. Ser. 65, Clouds, Cores, 
and Low Mass Stars, ed. D. P. Clemens \& R. Barvainis (San Fransisco: ASP), 87}
\reference{}{Crutcher, R. M. 1999, \apj, 520, 706}
\reference{}{Crutcher, R. M., Kaz\`es, I. 1983, \aap, 125, L23 (CK83)}
\reference{}{Crutcher, R. M., Kaz\`es, I., \& Troland, T. H. 1987,
\aap, 181, 119}
\reference{}{Crutcher, R. M., Roberts, D. A., Troland, T. H., \& Goss,
W. M. 1999a, \apj, 515, 275}
\reference{}{Crutcher, R. M., \& Troland, T. H. 2000, \apj, 537, L139}
\reference{}{Crutcher, R. M., Troland, T. H., Goodman, A. A., Heiles,
C., Kaz\`es, I., \& Myers, P. C. 1993, \apj, 407, 175}
\reference{}{Crutcher, R. M., \& Troland, T. H., Lazareff, B., \& Kaz\`es, I. 
1996, \apj, 456, 217}
\reference{}{Crutcher, R. M., \& Troland, T. H., Lazareff, B., Paubert,
G., \& Kaz\`es, I. 1999b, \apj, 514, L121}
\reference{}{Dame, T.M., Ungerechts, H., Cohen, R.S., De Geus, E.J.,
Grenier, I.A., May, J., Murphy, D.C., Nyman, L.-\AA., \& Thaddeus, P. 
1987, \apj, 322, 706}
\reference{}{De Lara, E., Chavarr\'{\i}a-K., C., \& L\'opez Molina, G. 1991, 
\aap, 243, 139}
\reference{}{Dickel, H. R., Dickel, J. R., \& Wilson, W. J. 1977, \apj,
217, 56}
\reference{}{Dickel, H. R., Seacord, A. W., II, \& Gottesman, S. T.
1977, \apj, 218, 133}
\reference{}{Dickel, H. R., \& Wall, J. V. 1974, \aap, 31, 5}
\reference{}{Elmegreen, B. G. 2000, \apj, 530, 277}
\reference{}{Frogel, J. A., \& Persson, S. E. 1974, \apj, 192, 351}
\reference{}{Goodman, A. A. 1989, PhD. Thesis, Harvard Univ.}
\reference{}{Goodman, A. A., Crutcher, R. M., Heiles, C., Myers, P. C.,
\& Troland, T. H. 1989, \apj, 338, L61}
\reference{}{Goodman, A. A., \& Heiles, C. 1994, \apj, 424, 208}
\reference{}{Goss, W. M. 1968, \apjs, 15, 131}
\reference{}{Goss, W. M., Manchester, R. N., \& Robinson, B. J. 1970,
Aust. J. Phys., 23, 559}
\reference{}{Hartmann, L. 2001, \aj, in press}
\reference{}{Hayakawa, T., Mizuno, A., Onishi, T., Yonekura, Y., Hara,
A., Yamaguchi, R., \& Fukui, Y. 1999, PASJ, 51, 919}
\reference{}{Heiles, C. 1988, \apj, 324, 321}
\reference{}{Heiles, C., Goodman, A. A., McKee, C. F., \& Zweibel, E.
G. 1993, in Protostars and Planets III, ed. E. H. Levy \& J. I.
Lunine (Tucson: Univ. Arizona Press), 279}
\reference{}{Hildebrand, R. H., Dotson, J., L., Dowell, C. D., Platt, S.
R., Schleuning, D., Davidson, J. A., \& Novak, G. 1995, in ASP Conf.
Ser. 73, Airborne Astronomy Symp. on the Galactic Ecosystem: From Gas
to Stars to Dust, ed. M. R. Haas, J. A. Davidson, \& E. F. Erikson (San
Fransisco: ASP), 97}
\reference{}{Jaffe, D. T., Stier, M. T., \& Fazio, G. G. 1982, \apj,
252, 601}
\reference{}{Kaz\`es, I., \& Crutcher, R. M. 1986, \aap, 164, 328}
\reference{}{Kaz\`es, I., Troland, T. H., Crutcher, R. M., \& Heiles,
C. 1988, \apj, 335, 263}
\reference{}{Kraemer, K. E. 1998, PhD Thesis, Boston Univ.}
\reference{}{Kraemer, K. E., \& Jackson, J. M. 1999, \apjs, 124, 439}
\reference{}{Lockman, F. J. 1989, \apjs, 71, 469}
\reference{}{Manchester, R. N., Robinson, B. J., \& Goss, W. M. 1970,
Aust. J. Phys., 23, 751}
\reference{}{Massi, F., Brand, J., \& Felli, M. 1997, \aap, 320, 972}
\reference{}{McGregor, P. J., Harrison, T. E., Hough, J. H., \& Bailey,
J. A. 1994, \mnras, 267, 755}
\reference{}{McKee, C. F. 1999, in The Origin of Stars \& Planetary Systems, 
ed. C. J. Lada, N. D. Kylafis (Dordrecht: Kluwer), 29}
\reference{}{McKee, C. F., Zweibel, E., Goodman, A. A., Heiles, C. 1993, 
in Protostars and Planets III, ed. E. H. Levy \& J. I.
Lunine (Tucson: Univ. Arizona Press), 327}
\reference{}{Mac Low M.-M., Klessen R. S., Burkert, A., \& Smith, M. D. 1998,
\prl, 80, 2754}
\reference{}{Mouschovias, T. Ch., \& Spitzer, L. 1976, \apj, 210, 326}
\reference{}{Myers, P. C., Ho, P. T. P., Schneps, M. H., Chin, G.,
Pankonin, V., \& Winnberg, A. 1978, \apj, 220, 864}
\reference{}{Nakano, T. 1998, \apj, 494, 587}
\reference{}{Nakano, T., \& Nakamura, T. 1978, PASJ, 30, 671}
\reference{}{Neckel, T. 1978, \aap, 69, 51}
\reference{}{Padoan, P, \& Nordlund, \AA. 1999, \apj, 526, 279}
\reference{}{Persi, P., Roth, M., Tapia, M., Ferrari-Toniolo, M., \&
Marenzi, A. R. 1994, \aap, 282, 474}
\reference{}{Pratap, P., Megeath, S. T., Bergin, E. A. 1999, \apj, 517, 799}
\reference{}{Reid, M. J., Myers, P. C., \& Bieging, J. H. 1987, \apj,
312, 830}
\reference{}{Reid, M. J., \& Silverstein, E. M. 1990, \apj, 361, 483}
\reference{}{Roberts, D. A., Crutcher, R. M., \& Troland, T. H. 1995,
\apj, 442, 208}
\reference{}{Roberts, D. A., Crutcher, R. M., Troland, T. H., \& Goss,
W. M. 1993, \apj, 412, 675}
\reference{}{Robinson, B. J., Goss, W. M., Manchester, R. N. 1970,
Aust. J. Phys., 23, 363}
\reference{}{Robinson, B. J., Caswell, J. L., Goss, W. M. 1971,
\aplett, 9, 5}
\reference{}{Rodr\'{\i}guez, L. F., Cant\'o, J, Moran, J. M. 1982, \apj,
255, 103}
\reference{}{Sarma, A. P., Troland, T. H., Roberts, D. A., \& Crutcher,
R. M. 2000, \apj, 533, 271}
\reference{}{Sault, R. J., Killeen, N. E. B., Zmuidzinas, J., \&
Loushin, R. 1990, \apjs, 74, 437}
\reference{}{Schwartz, R. D. 1991, in Loss Mass Star Formation in
Southern Molecular Clouds, ESO Scientific Report 11, ed. B. Reipurth
(Garching: ESO), 93}
\reference{}{Schwartz, P. R., Thronson, H. A. Jnr., Odenwald, S. F.,
Glaccum, W., Loewenstein, R. F, \& Wolf, G. 1985, \apj, 292, 231}
\reference{}{Shu, F. H., Allen, A., Shang, H., Ostriker, E. C., \& Li,
Z-Y. 1999, in The Origin of Stars \& Planetary Systems, ed. C. J. Lada,
N. D. Kylafis (Dordrecht: Kluwer), 193}
\reference{}{Smith, C. H., Bourke, T. L., Wright, C. M., Spoon, H. W. W.,
Aitken, D. K., Robinson, G., Storey, J. W. V., Fujiyoshi, T., 
Roche, P. F., Lehmann, T. 1999, \mnras, 303, 367}
\reference{}{Stone, J. M., Ostriker, E. C., \& Gammie, C. F. 1998, \apjl,
508, 99}
\reference{}{Tapia, M., Roth, M., Marraco, H., \& Ruiz, M. T. 1988,
\mnras, 232, 661}
\reference{}{Tomisaka, K., Ikeuchi, S., \& Nakamura, T. 1988, \apj, 335, 239}
\reference{}{Toriseva, M., H\"oglund, B., \& Mattila, K. 1985, 
Rev. Mexicana Astron. Af., 10, 135}
\reference{}{Troland, T. H. 1990, in Galactic and Intergalactic
Magnetic Fields, ed. R. Beck, P. P. Kronberg, \& R Wielebinski
(Dordrecht: Kluwer), 293}
\reference{}{Troland, T. H., Crutcher, R. M., \& Kaz\`es, I. 1986,
\apj, 304, L57}
\reference{}{Troland, T. H., \& Heiles, C. 1982, \apj, 252, 179}
\reference{}{Turner, B. E. 1973, \apj, 186, 357}
\reference{}{Turner, B. E. 1979, \aaps, 37, 1}
\reference{}{Turner, B. E., \& Heiles, C. 1971, \apj, 170, 453}
\reference{}{Whiteoak, J. B., \& Gardner, F. F. 1974, \aap, 37, 389}
\reference{}{Whittet, D. C. B., Prusti, T., Franco, G. A. P.,
Gerakines, P. A., Kilkenny, D., Larson, K. A., \& Wesselius, P. R.
1997, \aap, 327, 1194}
\reference{}{Wilking, B. A., Blackwell, J. H., Mundy, L. G., \& Howe, J.
E. 1989, \apj, 345, 257}
\reference{}{Wouterloot, J. G. A., \& Brand, J. 1989, \aaps, 80, 149}
\reference{}{Wouterloot, J. G. A., Brand, J., \& Fiegle, K. 1993, \aaps,
98, 589}
\reference{}{Wouterloot, J. G. A., \& Habing, H. J. 1985, \aaps, 60, 43}
\end{references}
\end{document}